%% file: CW_AMXP_O3_v7.tex
\documentclass[%
reprint,
superscriptaddress,
nofootinbib,
amsmath,amssymb,
aps,
prd,
]{revtex4-2}

\usepackage[table]{xcolor} % for red text colouring
\usepackage{booktabs}	% Better tables
\usepackage{array}		% More tables options
\usepackage[inline]{enumitem}	% Options for lists
\usepackage{multirow}   % Yet more tables options
\usepackage{dcolumn}% Align table columns on decimal point
\newcolumntype{.}{D{.}{.}{-1}}
\newcolumntype{d}[1]{D{.}{.}{#1}}
\usepackage{hyperref}
\hypersetup{
    colorlinks = true,
	citecolor = blue,
	linkcolor = blue
}
\usepackage{graphicx}
\usepackage{subcaption}
\usepackage{threeparttable}
\usepackage{pifont}
\usepackage{bbm}

\newcommand{\mj}{\mathcal{J}}
\newcommand{\mf}{\mathcal{F}}
\newcommand{\ml}{\mathcal{L}}
\newcommand{\lth}{\mathcal{L}_{\rm th}}
\newcommand{\lthg}{\mathcal{L}_{\rm th,\,G}}
\newcommand{\lthot}{\mathcal{L}_{\rm th,\,OT}}
\newcommand{\tdr}{T_{\textrm{drift}}}
\newcommand{\tasc}{T_{\textrm{asc}}}
\newcommand{\degr}{^{\circ}}
\newcommand{\pn}{p_{\rm noise}}
\newcommand{\hul}{h_0^{95\%}}
\newcommand{\hete}{HETE J1900.1$-$2455}
\newcommand{\igra}{IGR J00291$+$5934}
\newcommand{\igrb}{IGR J16597$-$3704}
\newcommand{\igrc}{IGR J17062$-$6143}
\newcommand{\igrd}{IGR J17379$-$3747}
\newcommand{\igre}{IGR J17498$-$2921}
\newcommand{\igrf}{IGR J17511$-$3057}
\newcommand{\igrg}{IGR J17591$-$2342}
\newcommand{\igrh}{IGR J18245$-$2452}
\newcommand{\igri}{IGR J17494$-$3030}
\newcommand{\maxi}{MAXI J0911$-$655}
\newcommand{\ngc}{NGC 6440 X$-$2}
\newcommand{\sax}{SAX J1808.4$-$3658}
\newcommand{\saxb}{SAX J1748.9$-$2021}
\newcommand{\swift}{Swift J1756.9$-$2508}
\newcommand{\swiftb}{Swift J1749.4$-$2807}
\newcommand{\xtea}{XTE J0929$-$314}
\newcommand{\xteb}{XTE J1751$-$305}
\newcommand{\xtec}{XTE J1807$-$294}
\newcommand{\xted}{XTE J1814$-$338}
\newcommand{\sco}{Scorpius X-1}

\begin{document}
\title{Search for continuous gravitational waves from 20 accreting millisecond X-ray pulsars in O3 LIGO data}

\input{authors.tex}

\collaboration{The LIGO Scientific Collaboration, the Virgo Collaboration, and the KAGRA Collaboration}

\begin{abstract}
Results are presented of searches for continuous gravitational waves from 20 accreting millisecond X-ray pulsars with accurately measured spin frequencies and orbital parameters, using data from the third observing run of the Advanced LIGO and Advanced Virgo detectors. The search algorithm uses a hidden Markov model, where the transition probabilities allow the frequency to wander according to an unbiased random walk, while the $\mj$-statistic maximum-likelihood matched filter tracks the binary orbital phase. Three narrow sub-bands are searched for each target, centered on harmonics of the measured spin frequency. The search yields 16 candidates, consistent with a false alarm probability of 30\% per sub-band and target searched. These candidates, along with one candidate from an additional target-of-opportunity search done for \sax, which was in outburst during one month of the observing run, cannot be confidently associated with a known noise source. Additional follow-up does not provide convincing evidence that any are a true astrophysical signal. When all candidates are assumed non-astrophysical, upper limits are set on the maximum wave strain detectable at 95\% confidence, $\hul$. The strictest constraint is $\hul = 4.7\times 10^{-26}$ from \igrc. Constraints on the detectable wave strain from each target lead to constraints on neutron star ellipticity and $r$-mode amplitude, the strictest of which are $\epsilon^{95\%} = 3.1\times 10^{-7}$ and $\alpha^{95\%} = 1.8\times 10^{-5}$ respectively. This analysis is the most comprehensive and sensitive search of continuous gravitational waves from accreting millisecond X-ray pulsars to date.
\end{abstract}


\maketitle

\section{\label{sec:intro}Introduction} 
Second generation, ground-based gravitational wave detectors, specifically the Advanced Laser Interferometer Gravitational wave Observatory (Advanced LIGO) \cite{aligo2015} and Advanced Virgo \cite{avirgo2014}, have detected more than 50 compact binary coalescence events in recent years \cite{gwtc1, gwtc2, gwtc2.1}. Continuous gravitational waves from rapidly-rotating neutron stars are also potential sources, e.g.~a non-axisymmetry due to mountains on the surface, or stellar oscillation modes in the interior \cite{Glampedakis2018, Sieniawska2019, Haskell2021}. There are no reported detections of continuous gravitational waves to date, despite a number of searches in Advanced LIGO and Advanced Virgo data \cite{cw170817, o2narrow, o2vitsco, Middleton2020, Papa2020, Fesik2020, Fesik2020a, Piccinni2020, Steltner2021, Zhang2021, Beniwal2021, Jones2021, o3aknown, Dergachev2021a, Rajbhandari2021, o3abinaryallsky, Wette2021, o30537_2f, o30537_rmodes, o3aAllSkyIso, o3aSNR, Ashok2021}.

Low-mass X-ray binaries (LMXBs) are a high-priority target for continuous gravitational wave searches. LMXBs are composed of a compact object, such as a neutron star\footnote{LMXBs in which the compact object is a stellar-mass black hole are not expected to function as continuous gravitational wave sources and are not discussed in this paper.}, which accretes matter from a stellar-mass ($\lesssim 1 M_\odot$) companion \cite{Patruno2021}. The accretion exerts a torque that may spin up the compact object. Electromagnetic (EM) observations show that even the pulsar with the highest known frequency, PSR J1748$-$2446ad at 716\,Hz \cite{Hessels2006}, rotates well below the centrifugal break-up frequency, estimated at $\sim1400\,$Hz \cite{Cook1994}. Gravitational wave emission may provide the balancing torque in binary systems such as these, stopping the neutron star from spinning up to the break-up frequency \cite{Bildsten1998, Andersson1999a}. If so, there should thus be a correlation between accretion rate (which is inferred via X-ray flux) and the strength of the continuous gravitational wave emission \cite{Papaloizou1978, Wagoner1984, Bildsten1998, Andersson1999a}. The LMXB \sco\ is the brightest extra-Solar X-ray source in the sky, making it a prime target for searches for continuous gravitational waves \cite{o1vitsco, o1crosscorSco, o2vitsco, Zhang2021}. 

Some LMXBs have EM observations of pulsations during ``outburst'' events lasting days to months, which allow for measurement of their rotational frequency, $f_\star$, to an accuracy of $\sim10^{-8}\,$Hz, and measurement of their binary ephemerides \cite{DiSalvo2020, Patruno2021}. LMXBs that are observed to go into outburst and have measurable pulsations with millisecond periods are sometimes called accreting millisecond X-ray pulsars (AMXPs). If the rotational frequency is known, computationally cheap narrowband searches are possible. Six AMXPs were previously searched for continuous gravitational waves, one in Science Run 6 (S6) using the TwoSpect algorithm \cite{twoSpectInit,s6twoSpectScoXTE}, and five in Observing Run 2 (O2) using the same Hidden Markov Model (HMM) algorithm we use in this work \cite{Suvorova2017,Middleton2020}. No significant candidates were found in either search. Searches for continuous gravitational waves from LMXBs are difficult as the rotation frequency may wander stochastically on timescales of $\lesssim 1\,$yr \cite{Mukherjee2018}, limiting the duration of coherent integration. A HMM tracks a wandering signal, and is the search algorithm we use here, following Refs.~\cite{Suvorova2016,Suvorova2017,o2vitsco,Middleton2020}.

Advanced LIGO and Advanced Virgo began the third Observing Run (O3) on April 1 2019, 15:00 UTC. There was a month-long commissioning break between October 1 2019, 15:00 UTC, and November 1 2019, 15:00 UTC, after which observations resumed until March 27, 2020, 17:00 UTC. This month-long break divides O3 into two segments: O3a and O3b. In this work we search the full O3 data set for continuous gravitational wave signals from AMXPs with known rotational frequencies. The search is a more sensitive version of an analogous search in O2 data \cite{Middleton2020}, with an expanded target list. We briefly review the algorithm and O2 search in Sec.~\ref{sec:alg}. In Secs.~\ref{sec:targets} and \ref{sec:params} we describe the targets and the parameter space respectively. We discuss the data used in Sec.~\ref{sec:o3data}. In Sec.~\ref{sec:vetoes} we describe the vetoes applied to discriminate between terrestrial and astrophysical candidates. In Sec.~\ref{sec:o3results} we present the results of the search. In Sec.~\ref{sec:shortsax} we describe an additional target-of-opportunity search performed for one of the targets that was in outburst during O3a. We provide upper limits for the detectable wave-strain, and astrophysical implications thereof, in Sec.~\ref{sec:ul}. We conclude in Sec.~\ref{sec:concl}.

\section{Search algorithm \label{sec:alg}}
The search in this paper follows the same prescription as the O2 searches for \sco\ \cite{o2vitsco} and LMXBs with known rotational frequency \cite{Middleton2020}. It is composed of two parts: a HMM which uses the Viterbi algorithm to efficiently track the most likely spin history, and the $\mj$-statistic, which calculates the likelihood a gravitational wave is present given the detector data, and the orbital parameters of both the Earth and the LMXB. The HMM formalism is identical to that used in Refs.~\cite{Suvorova2016, o1vitsco, Suvorova2017, o2vitsco, Middleton2020}, and the $\mj$-statistic was first introduced in Ref.~\cite{Suvorova2017}. Below, we provide a brief review of both the HMM and the $\mj$-statistic.
 
\subsection{HMM\label{sec:hmm}} 
In a Markov process, the probability of finding the system in the current state depends only on the previous state. In a hidden Markov process the states are not directly observable and must be inferred from noisy data. In this paper, the hidden state of interest is the gravitational wave frequency $f(t)$. Although the rotation frequency $f_\star(t)$ of every target in this search is measured accurately from EM pulsations, we allow $f(t) \neq f_\star(t)$ in general for three reasons:
\begin{enumerate*}[label=\roman*)] \item different emission mechanisms emit at different multiples of $f_\star$ \cite{Riles2013a};
\item a small, fluctuating drift may arise between $f(t)$ and $f_\star(t)$, if the star's core (where the gravitational-wave-emitting mass or current quadrupole may reside) decouples partially from the crust (to which EM pulsations are locked) \cite{Suvorova2016, s5crab}; and,
\item the rotational frequency of the crust may also drift stochastically due to a fluctuating accretion torque \cite{Mukherjee2018, Patruno2021}.
\end{enumerate*}
The gravitational-wave frequency is therefore hidden even though the EM measurement of $f_\star$ helps restrict the searched frequency space, as described in Sec.~\ref{sec:params}.

Following the notation of Refs.~\cite{o2vitsco, Middleton2020} we label the hidden state variable as $q(t)$. In our model, it transitions between a discrete set of allowed values $\{q_1,...,q_{N_Q}\}$ at discrete times $\{t_0,...,t_{N_T}\}$. The probability of the state transitioning from $q_i$ at time $t_n$ to $q_j$ at time $t_{n+1}$ is determined by the transition matrix $A_{q_j q_i}$. In this search, as in previous searches of LMXBs \cite{o1vitsco, o2vitsco, Middleton2020}, the transition matrix is
\begin{equation}
\label{eq:transmat}
{A_{q_j q_i} = \frac{1}{3}\left(\delta_{q_j q_{i+1}} + \delta_{q_j q_i} + \delta_{q_j q_{i-1}}\right)}\ ,
\end{equation}
where $\delta_{ij}$ is the Kronecker delta. Eq.~\eqref{eq:transmat} corresponds to allowing $f(t)$ to move 0, or $\pm1$ frequency bins, with equal probability, at each discrete transition. It implicitly defines the signal model for $f(t)$ to be a piece-wise constant function, with jumps in frequency allowed at the discrete times $\{t_0, ... , t_{N_T}\}$. This is a well-tested approximation for an unbiased random walk \cite{Suvorova2016, Suvorova2017}.

The total duration of the search is $T_\textrm{obs}$, which we split into $N_T$ coherent equal chunks of length $\tdr$, where $N_T = \lfloor T_\textrm{obs} / \tdr \rfloor$, and $\lfloor ... \rfloor$ indicates rounding down to the nearest integer. We justify our choice of $\tdr$ in Sec.~\ref{sec:params}. In essence, it needs to be short enough to ensure that $f_\star(t)$ does not wander by more than one frequency bin during each time segment, but ideally no shorter in order to maximize the signal-to-noise ratio in each segment. For each time segment the likelihood that the observation $o_j$ is related to the hidden state $q_i$ is given by the emission matrix $L_{o_j q_i}$. We calculate $L_{o_j q_i}$ from the data via a frequency domain estimator, e.g.~the $\mj$-statistic, as discussed in Sec.~\ref{sec:jstat}.

The probability that the hidden path is $Q = \{q(t_0), ..., q(t_{N_T})\}$ given a set of observations $O = \{o(t_0), ..., o(t_{N_T})\}$ is
\begin{align}
\label{eq:probq_o}
P(Q\,|\,O) =&~ \Pi_{q(t_0)}\, A_{q(t_1) q(t_0)} L_{o(t_1) q(t_1)}\ ... \nonumber\\
&~\times  A_{q(t_{N_T}) q(t_{N_T - 1})} L_{o(t_{N_T}) q(t_{N_T})}\ ,
\end{align}
where $\Pi_{q(t_0)}$ is the prior probability of starting in the state $q(t_0)$, and is taken to be uniform within a certain range guided by EM measurements of $f_\star$. The Viterbi algorithm is a computationally efficient way to find the path $Q^{*}$ that maximizes Eq.~\eqref{eq:probq_o} \cite{Viterbi1967}. 

The detection statistic we use in this work is $\ml = \ln P(Q^{*}\,|\,O)$, i.e.~the log-likelihood of the most likely path given the data. The search outputs one $P(Q^{*}|O)$ value per frequency bin, corresponding to the optimal path $Q^{*}$ terminating in that frequency bin.
   
\subsection{$\mj$-statistic\label{sec:jstat}} 
Any long-lived gravitational wave signal from an LMXB observed by the detectors is Doppler modulated by the orbital motion of the detectors around the Solar System barycenter, and by the orbital motion of the compact object in its binary. The $\mf$-statistic is a frequency domain estimator originally designed for isolated neutron stars, and accounts for the Earth's annual orbital motion (as well as the amplitude modulation caused by the Earth's diurnal rotation) \cite{Jaranowski1998}. Algorithms that implement the $\mf$-statistic, such as \texttt{lalapps\_ComputeFstatistic\_v2} \cite{LAL2018}, have subsequently added functionality to account for modulation of the signal due to binary motion. 

The $\mj$-statistic accounts for the binary modulation via a Jacobi-Anger expansion of the orbit \cite{Suvorova2017}. It ingests $\mf$-statistic ``atoms'' as calculated for an isolated source as an input, assumes the binary is in a circular orbit\footnote{This assumption is justified as none of the targets described in Sec.~\ref{sec:targets} have measurable eccentricity with sufficient precision \cite{DiSalvo2020, Patruno2021}.}, and requires three binary orbital parameters: the period $P$, the projected semi-major axis $a_0$, and the time of passage of the ascending node $\tasc$. We use the $\mj$-statistic as the frequency domain estimator $L_{o_j q_i}$ in this paper, as in Refs.~\cite{o2vitsco, Middleton2020}. The $\mj$-statistic is a computationally efficient algorithm, as it re-uses $\mf$-statistic atoms when searching over a template bank of binary orbital parameters. 

\section{\label{sec:targets}Targets} 
The AMXPs chosen as targets for this search, along with their positions, orbital elements, and pulsation frequencies are listed in Table \ref{tab:info}. These 20 targets constitute all known AMXPs with observed coherent pulsations and precisely measured orbital elements as of April 2021\footnote{We do not include the AMXP Aquila X-1 \cite{Casella2008, MataSanchez2017} in our target list as there is a large uncertainty on all three binary orbital elements, compared to the other 20 AMXPs. One would need to search $>10^{10}$ binary orbital templates, an order of magnitude more than the rest of the targets combined. The number of binary orbital templates is calculated as a function of the uncertainty in orbital elements in Sec.~\ref{sec:numtemps}.}. For details on the relevant EM observations, principally in the X-ray band, see Refs.~\cite{Watts2008, Marino2019, DiSalvo2020, Patruno2021}.

Most AMXPs are transient, with ``active'' (outburst) and ``quiescent'' phases. Pulsations, and therefore $f_\star$, are only observed during the active phase. Active phases are typically associated with accretion onto the neutron star, however accretion can also happen during quiescence \cite{Melatos2016}. The frequency derivatives, $\dot{f}_\star$, in the active phase and in the quiescent phase are set by the accretion torque and magnetic dipole braking respectively \cite{Ghosh1977, Melatos2016}. The value of $\dot{f}_\star$ has implications for the continuous gravitational wave signal strength (see Sec.~\ref{sec:ul_comp}), as well as the choice of $\tdr$ (see Sec.~\ref{sec:tdr}).

One target, \sax, went into outburst during O3a \cite{Bult2019, Bult2020, Goodwin2020}. It may be the case that continuous gravitational waves are only emitted when an AMXP is in outburst \cite{Haskell2017a}. If so, we increase our signal-to-noise ratio by searching only data from the times that it was in outburst, compared to searching the entirety of O3 data. To investigate this possibility, we perform in Sec.~\ref{sec:shortsax} an additional target-of-opportunity search for continuous gravitational waves from \sax\ while it is in outburst.

\begin{turnpage}
\begin{table*}
\begin{threeparttable}
\caption{Target list: position (RA and Dec), orbital period ($P$), projected semi-major axis in light-seconds ($a_0$), time of passage through the ascending node as measured near the time of the most recent outburst ($\tasc$), the time of passage through the ascending node as propagated to the start of O3 ($T_\textrm{asc,\,O3}$), as described in Sec.~\ref{sec:numtemps}, and frequency of observed pulsations ($f_\star$). Numbers in parentheses indicate reported 1$\sigma$ errors (68\% confidence level), unless otherwise noted. All objects have positional uncertainty $\leq1$s in RA and $\leq0.5''$ in Dec.} \label{tab:info}
{\renewcommand{\arraystretch}{1.3}% for the vertical padding
\begin{ruledtabular}
\begin{tabular}{l l l l l l l l l}
Target & RA & Dec & P$/$s & $a_0/$lt-s & $\tasc/$GPS time & $T_{\textrm{asc,\,O3}}/$GPS time &   $f_\star / $Hz & Refs.~\\
\midrule
IGR J00291$+$5934    & 00h29m03.05s  & $+59\degr34'18.93''$    & 8844.07673(9) & 0.064993(2)   & 1122149932.93(5)                                       & 1238157687(1)     & 598.89213099(6)  & \cite{Torres2008,Patruno0029} \\
MAXI J0911$-$655     & 09h12m02.46s  & $-64\degr52'06.37''$    & 2659.93312(47)& 0.017595(9)   & 1145507148.0(9)                                        & 1238165918(16)      & 339.9750123(3)     & \cite{Homan2016, Sanna0911}  \\
XTE J0929$-$314      & 09h29m20.19s  & $-31\degr23'03.2''$     & 2614.746(3)   & 0.006290(9)   & 705152406.1(9)                                         & 1238165763(612)      & 185.105254297(9) & \cite{Galloway2002, Giles2005}  \\
IGR J16597$-$3704    & 16h59m32.902s & $-37\degr07'14.3''$     & 2758.2(3)     & 0.00480(3)    & 1193053416(9)                                          & 1238163777(4907)      & 105.1758271(3)   & \cite{Tetarenko2018, Sanna16597} \\
IGR J17062$-$6143    & 17h06m16.29s  & $-61\degr42'40.6''$     & 2278.21124(2) & 0.003963(6)   & 1239389342(4)                                          & 1238165942(4)     & 163.656110049(9) & \cite{Bult2021}  \\
IGR J17379$-$3747    & 17h37m58.836s & $-37\degr46'18.35''$    & 6765.8388(17) & 0.076979(14)  & 1206573046.6(3)                                        & 1238162748(8)     & 468.083266605(7) & \cite{Sanna17379, Bult17379}  \\
SAX J1748.9$-$2021   & 17h48m52.161s & $-20\degr21'32.406''$   & 31555.300(3)  & 0.38757(2)    & 1109500772.5(8)                                        & 1238151731(12)      & 442.3610957(2)   & \cite{Sanna1748, DiSalvo2020} \\
NGC 6440 X$-$2       & 17h48m52.76s  & $-20\degr21'24.0''$     & 3457.8929(7)  & 0.00614(1)    & 956797704(2)                                           & 1238166449(57)      & 205.89221(2)     & \cite{Heinke2010a, Bult2015}  \\
IGR J17494$-$3030	 & 17h49m23.62s  & $-30\degr29'58.999''$   & 4496.67(3)    & 0.015186(12)  & 1287797911(1)                                          & 1238163668(331)      & 376.05017022(4)  & \cite{Ng2021}   \\
Swift J1749.4$-$2807 & 17h49m31.728s & $-28\degr08'05.064''$   & 31740.8417(27)& 1.899568(11)  & 1298634645.85(12)                                      & 1238136602(5)     & 517.92001385(6) & \cite{Jonker2013, Bult2021a, Sanna1749} \\
IGR J17498$-$2921    & 17h49m56.02s  & $-29\degr19'20.7''$     & 13835.619(1)\tnote{b} & 0.365165(5)\tnote{b}  & 997147537.43(7)\tnote{b}               & 1238164020(6)     & 400.99018734(9)\tnote{b} & \cite{Papitto17498, Falanga2012}  \\
IGR J17511$-$3057    & 17h51m08.66s  & $-30\degr57'41.0''$     & 12487.5121(4) & 0.2751952(18) & 936924316.03(3)                                        & 1238160570(10)      & 244.83395145(9)  & \cite{Paizis2012, Riggio17511} \\
XTE J1751$-$305      & 17h51m13.49s  & $-30\degr37'23.4''$     & 2545.3414(38)\tnote{a} & 0.010125(5)\tnote{a} & 701914663.57(3)\tnote{a}				& 1238164644(487)       & 435.31799357(3)\tnote{a}  & \cite{Markwardt1751, Papitto1751} \\
Swift J1756.9$-$2508 & 17h56m57.43s  & $-25\degr06'27.4''$     & 3282.40(4)    & 0.00596(2)    & 1207196675(9)                                          & 1238166119(378)      & 182.06580377(11) & \cite{Sanna1756}  \\
IGR J17591$-$2342    & 17h59m02.86s  & $-23\degr43'08.3''$     & 31684.7503(5) & 1.227714(4)   & 1218341207.72(8)                                       & 1238144176.7(3)          & 527.425700578(9) & \cite{Russell2018, Sanna17591} \\
XTE J1807$-$294      & 18h06m59.8s   & $-29\degr24'30''$       & 2404.4163(3)  & 0.004830(3)   & 732384720.7(3)                                         & 1238165711(63)      & 190.62350702(4)  & \cite{Markwardt1807, Patruno1807}  \\
SAX J1808.4$-$3658   & 18h08m27.647s & $-36\degr58'43.90''$    & 7249.155(3)   & 0.062809(7)   & 1250296258.5(2)                                        & 1238161173(5)     & 400.97521037(1)  & \cite{Bult2020} \\
XTE J1814$-$338      & 18h13m39.02s  & $-33\degr46'22.3''$     & 15388.7229(2)\tnote{a}\hspace{0.5em} & 0.390633(9)\tnote{a} & 739049147.41(8)\tnote{a} & 1238151597(4)     & 314.35610879(1)\tnote{a} & \cite{Krauss2005, Papitto1814}  \\
IGR J18245$-$2452    & 18h24m32.51s  & $-24\degr52'07.9''$     & 39692.812(7)  & 0.76591(1)    & 1049865088.37(9)                                       & 1238128096(33)      & 254.3330310(1)   & \cite{Pallanca2013, Papitto18245}  \\
HETE J1900.1$-$2455  & 19h00m08.65s  & $-24\degr55'13.7''$     & 4995.2630(5)  & 0.01844(2)    & 803963262.3(8)                                         & 1238161513(43)      & 377.296171971(5) & \cite{Fox2005, Kaaret2006, Patruno1900} \\
\end{tabular}
\end{ruledtabular}
}
\begin{tablenotes}
\item[a]{90\% confidence level}
\item[b]{3$\sigma$ error}
\end{tablenotes}
\end{threeparttable}
\end{table*}
\end{turnpage}

\section{Search parameters \label{sec:params}} 
The $\mj$-statistic matched filter requires specification of the source sky position [right ascension (RA) and declination (Dec)], the orbital period $P$, the projected semi-major axis $a_0$, and the orbital phase $\phi_a$ at the start of the search. The orbital phase can be equivalently specified via a time of passage through the ascending node, $\tasc$. EM observations constrain all of these parameters, as well as the spin frequency $f_\star$. These measurements, along with their associated uncertainties, are listed in Table \ref{tab:info}.

There are several mechanisms that could lead to continuous gravitational wave emission from an AMXP, in its active or quiescent phase. ``Mountains'' on the neutron star surface, be they magnetically or elastically supported, emit at 2$f_\star$ and potentially $f_\star$ \cite{Jones2010}. The dominant continuous gravitational wave emission from $r$-mode oscillations (Rossby waves excited by radiation-reaction instabilities) is predicted to be at $\sim4f_\star / 3$ \cite{Andersson1998, Friedman1998, Yoshida2001, Alford2012}. Thus, we search frequency sub-bands centered on $\{1, 4/3, 2 \}\,f_\star$ for each target. As in Refs.~\cite{o2vitsco, Middleton2020} we choose a sub-band width of $\sim0.61\,$Hz\footnote{Other narrowband searches, such as Refs.~\cite{o1narrow, o2narrow}, search sub-bands whose width, $\sim 10^{-3} f$, scales with frequency. We note that $0.61\,$Hz is comparable to $10^{-3}f$ for the harmonics of $f_\star$ that we search in this paper, but is $2^{20} \Delta f$, where $\Delta f$ is the frequency bin size defined in Sec.~\ref{sec:tdr}. Having the number of frequency bins in the sub-band equal a power of two speeds up the Fourier transform \cite{o2vitsco}.}.

Recent work indicates that the continuous gravitational wave signal from $r$-modes could emit at a frequency far from $4f_\star / 3$ due to equation-of-state-dependent relativistic corrections, and so comprehensive searches for $r$-modes may need to cover hundreds of Hz for the targets listed in Table \ref{tab:info} \cite{Idrisy2015, Caride2019}. The exact range of frequencies to search is a non-linear function of $f_\star$, and does not necessarily include $4f_\star /3$ (see equation (17) of Ref.~\cite{Caride2019}). However, these estimates are still uncertain. We deliberately search $\sim0.61\,$Hz sub-bands centered on $4f_\star / 3$, as an exhaustive broadband search lies outside the scope of this paper, which aims to conduct fast, narrowband searches at astrophysically motivated harmonics of $f_\star$ while accommodating frequency wandering within those sub-bands, a challenge in its own right.

\subsection{$\tdr$ and frequency binning \label{sec:tdr}}
Another key parameter for the search algorithm described in Sec.~\ref{sec:alg} is the coherence time $\tdr$. As in Refs.~\cite{o2vitsco, Middleton2020} we fix $\tdr=10\,$d for each target\footnote{We consider additional $\tdr$ durations for the target-of-opportunity search for continuous gravitational waves from \sax\ during its O3a outburst in Sec.~\ref{sec:shortsax}.}. This choice of $\tdr$ is guided by observations of \sco\ \cite{Mukherjee2018}. Quantitative studies of how X-ray flux variability in AMXPs impacts searches for continuous gravitational waves are absent from the literature. The choice to use $\tdr=10\,$d balances the increased sensitivity achieved via longer coherence times with the knowledge that the gravitational wave frequency may wander stochastically, e.g.~due to fluctuations in the mass accretion rate. The particular value $\tdr=10\,$d has been adopted in all previous Viterbi LMXB searches \cite{o1vitsco, o2vitsco, Middleton2020} and is justified approximately with reference to a simple random-walk interpretation of fluctuations in the X-ray flux of \sco\ \cite{Sammut2014, Messenger2015a, Mukherjee2018}, but other values are reasonable too. 

We remind the reader that the choice of $\tdr$ implicitly fixes the proposed signal model as one in which the frequency may wander step-wise zero, plus or minus one frequency bin every $\tdr=10\,$d. The size of the frequency bins, $\Delta f$, is fixed by the resolution implied by the coherence time, i.e.~$\Delta f = 1 / (2\, \tdr) = 5.787037\times10^{-7}\,$Hz, for $\tdr=10\,$d. As $\Delta f$ depends on $\tdr$, changing the coherence time explicitly changes the signal model, e.g.~if $\tdr$ is halved and $T_{\rm obs}$ is kept constant, then both $N_T$ and $\Delta f$ double; thus the signal can move up to a factor of four more in frequency in the same $T_{\rm obs}$. The connection between the coherence time and signal model features in all semi-coherent search methods. However, for a HMM-based search such as this, the choice of coherence time is not limited by computational cost, as it is in all-sky searches or searches based on the $\mathcal{F}$-statistic \cite{o2allsky, Steltner2021}.

This analysis does not search over any frequency derivatives. The maximum absolute frequency derivative, $|\dot{f}_{\rm max}|$, that does not change the frequency more than one frequency bin over the course of one coherent chunk is
\begin{equation}
\left|\dot{f}_{\rm max} \right| = \frac{\Delta f}{\tdr} \approx 6.7\times 10^{-13}\,\textrm{Hz\,s}^{-1}\ .
\end{equation}
When measured, the long-term secular frequency derivative is well below this value for all of our targets, see Sec.~\ref{sec:ul_comp} for details.

\subsection{Number of orbital templates \label{sec:numtemps}}
The orbital elements are known to high precision, with the uncertainty in $P$ satisfying $\sigma_P \lesssim 10^{-3}\,$s, the uncertainty in $a_0$ satisfying $\sigma_{a_0} \lesssim 10^{-4}\,$light-seconds (lt-s), and the uncertainty in $\tasc$ satisfying $\sigma_{\tasc} \lesssim 1\,$s. However, $\tasc$ is measured relative to the target's most recent outburst, which is often years before the start of O3 ($T_{\rm O3,\,start}=1238166483\,$GPS time). We need to propagate it forward in time. This propagation compounds the uncertainty in $\tasc$, viz.~ \cite{o1crosscorSco, o2vitsco, Middleton2020}
\begin{equation}
\sigma_{T_{\rm asc,\,O3}} = \left[\sigma_{\tasc}^2 + \left(N_{\rm orb} \sigma_P\right)^2\right]^{1/2}\,,
\end{equation}
where $N_{\rm orb}$ is the number of orbits between the observed $\tasc$ and $T_{\rm asc,\,O3}$. Henceforth $\tasc$ and $\sigma_{\tasc}$ symbolize their values when propagated to $T_{\rm O3,\,start}$.

To conduct the search over the orbital elements for each target and sub-band we construct a rectangular grid in the parameter space defined by $(P \pm 3\sigma_P, a_0 \pm 3\sigma_{a_0}, \tasc \pm 3\sigma_{\tasc})$. For three targets, \xtea, \igrb, and \igri, the range $(\tasc \pm P/2)$ is smaller than $(\tasc \pm 3\sigma_{\tasc})$ and we use the former. We assume that $P$ and $a_0$ remain within the same bin for the entire search. While some targets have a non-zero measurement of $\dot{P}\, T_{\rm obs}$ ($\dot{a}_0\, T_{\rm obs}$), in all cases it is much smaller than the template spacing in $P$ ($a_0$) \cite{Patruno1808, Patruno0029, Bult2021}.

It is unlikely that the true source parameters lie exactly on a grid point in the parameter space. Thus the grid is spaced such that the maximum mismatch, $\mu_{\rm max}$, is never more than an acceptable level. The mismatch is defined as the fractional loss in signal-to-noise ratio between the search executed at the true parameters and at the nearest grid point \cite{Leaci2015}. We calculate the number of grid points required for $P$, $a_0$ and $\tasc$ using Eq.~(71) of Ref.~\cite{Leaci2015}, i.e.
\begin{align}
N_P &= \pi^2\sqrt{6}\, \mu_{\rm max}^{-1/2} f a_0  \frac{\gamma \tdr}{P^2} \sigma_P\,,\label{eq:np}\\
N_{a_0} &= 3\pi\sqrt{2}\, \mu_{\rm max}^{-1/2} f \sigma_{a_0}	\,,\label{eq:na0}\\
N_{\tasc} &= 6\pi^2\sqrt{2}\, \mu_{\rm max}^{-1/2} f a_0 \frac{1}{P} \sigma_{\tasc}	\,,\label{eq:ntasc}
\end{align}
where $\gamma$ is a refinement factor defined in general in Eq.~(67) of Ref.~\cite{Leaci2015}. In the case of O3, the semi-coherent segments are contiguous so we have $\gamma = N_T = 36$. We fix $\mu_{\rm max} = 0.1$. A set of software injections into O3 data verifies that a template grid constructed with $\mu_{\rm max} = 0.1$ results in a maximum fractional loss in signal-to-noise ratio of 10\%. We make the conservative choice of rounding $N_P$, $N_{a_0}$, and $N_{\tasc}$ up to the nearest integer, after setting $f$ to the highest frequency in each $0.61\,$Hz sub-band. As in Ref.~\cite{Middleton2020} we find $N_{a_0} = 1$ for each target and sub-band, and so hold $a_0$ constant at its central value while searching over $P$ and $\tasc$. Table \ref{tab:ntemps} shows $N_P$, $N_{\tasc}$, and $N_{\rm tot} = N_P N_{\tasc}$ for each target and sub-band. When Eq.~\eqref{eq:np} or \eqref{eq:ntasc} predicts only two templates for a given sub-band we round up to three, ensuring that the central value of $P$ or $\tasc$ from EM observations is included in the template bank. Note that the EM observations are sufficiently precise that $<5\times10^{4}$ templates are required across all targets and sub-bands. This is in contrast to the O2 search for continuous gravitational waves from \sco, for which $\sim10^9$ templates were needed, mainly due to the large uncertainty in $a_0$, and the unknown rotation frequency \cite{o2vitsco}.

\begin{table*}
\caption{Starting frequencies, $f_{\rm s}$, for each $\sim0.61\,$Hz-wide sub-band, number of templates needed to cover the $P$ and $\tasc$ domains in that sub-band, $N_P$ and $N_{\tasc}$ respectively, and the total number of templates for each sub-band, $N_{\rm tot} = N_P N_{\tasc}$. The projected semi-major axis $a_0$ is known precisely enough that we have $N_{a_0}=1$ for each sub-band. \label{tab:ntemps}}
\begin{ruledtabular}
\begin{tabular}{l d{3} d{1.5} d{1.5} d{1.5} l d{3} d{1.5} d{1.5} d{1.5}}
\textrm{Target}	& \textrm{$f_{\rm s}$ (Hz)}  & \textrm{$N_P$} & \textrm{$N_{\tasc}$} & \textrm{$N_{\rm tot}$} & \textrm{Target}	& \textrm{$f_{\rm s}$ (Hz)}   & \textrm{$N_P$}     & \textrm{$N_{\tasc}$} & \textrm{$N_{\rm tot}$} \\
\midrule
IGR J00291$+$5934 	 	& 598.6 & 1 & 3 & 3 			& IGR J17498$-$2921 	& 400.7 & 1 & 17 & 17 			\\
		 				& 798.5 & 1 & 3 & 3 			& 						& 534.7 & 1 & 22 & 22 			\\
		 				& 1197.8 & 1 & 3 & 3 			& 						& 802.0 & 3 & 33 & 99 			\\
MAXI J0911$-$655 		& 339.7 & 1 & 10 & 10			& IGR J17511$-$3057 	& 244.5 & 1 & 14 & 14 			\\
      				 	& 453.3 & 3 & 14 & 42			& 						& 326.4 & 1 & 19 & 19 			\\
      				 	& 679.9 & 3 & 20 & 60			& 						& 489.7 & 1 & 28 & 28 			\\
XTE J0929$-$314 		& 184.8 & 3 & 52 & 156 			& XTE J1751$-$305 		& 435.0 & 4 & 195 & 780 		\\
    				   	& 246.8 & 3 & 69 & 207 			&      				  	& 580.4 & 5 & 260 & 1300 		\\
    				  	& 370.2 & 3 & 104 & 312 		&      				  	& 870.6 & 8 & 390 & 3120 		\\
IGR J16597$-$3704 	 	& 104.9 & 49 & 23 & 1127 		& Swift J1756.9$-$2508 	& 181.8 & 10 & 34 & 340 		\\
		 				& 140.2 & 65 & 31 & 2015 		&      				  	& 242.8 & 13 & 45 & 585 		\\
		 				& 210.4 & 97 & 46 & 4462 		&      				  	& 364.1 & 20 & 67 & 1340 		\\
IGR J17062$-$6143 	 	& 163.4 & 1 & 1 & 1 			& IGR J17591$-$2342 	& 527.1 & 1 & 3 & 3 			\\
		 				& 218.2 & 1 & 1 & 1 			& 						& 703.2 & 3 & 3 & 9 			\\
		 				& 327.3 & 1 & 1 & 1 			& 						& 1054.9 & 3 & 4 & 12 			\\
IGR J17379$-$3747 	 	& 467.8 & 4 & 12 & 48 			& XTE J1807$-$294 		& 190.3 & 1 & 7 & 7 			\\
						& 624.1 & 5 & 15 & 75 			&     				   	& 254.2 & 1 & 9 & 9 			\\
						& 936.2 & 7 & 23 & 161 			&     				   	& 381.2 & 1 & 13 & 13 			\\
SAX J1748.9$-$2021 		& 442.1 & 3 & 18 & 54 			& SAX J1808.4$-$3658 	& 400.7 & 4 & 5 & 20 			\\
						& 589.8 & 3 & 24 & 72 			&     					& 534.6 & 5 & 7 & 35 			\\
						& 884.7 & 3 & 36 & 108 			&     					& 802.0 & 7 & 10 & 70 			\\
NGC 6440 X$-$2 	 		& 205.6 & 1 & 6 & 6 			& XTE J1814$-$338 		& 314.1 & 1 & 9 & 9 			\\
       				  	& 274.5 & 1 & 8 & 8 			&     				    & 419.1 & 1 & 12 & 12 		    \\
       				 	& 411.8 & 1 & 12 & 12 			&     				   	& 628.7 & 1 & 17 & 17 			\\
IGR J17494$-$3030 	 	& 375.7 & 21 & 112 & 2352 		& IGR J18245$-$2452 	& 254.0 & 3 & 44 & 132 			\\
						& 501.4 & 27 & 150 & 4050 		& 						& 339.1 & 3 & 58 & 174 			\\
						& 752.1 & 41 & 224 & 9184 		& 						& 508.7 & 5 & 87 & 435 			\\
Swift J1749.4$-$2807 	& 517.6 & 7 & 43 & 301 			& HETE J1900.1$-$2455 	& 377.0 & 1 & 17 & 17 			\\
						& 690.6 & 9 & 57 & 513 			& 		 				& 503.1 & 1 & 22 & 22 			\\
						& 1035.8 & 13 & 85 & 1105 		& 		 	   			& 754.6 & 1 & 33 & 33 			\\
\end{tabular}
\end{ruledtabular}
\end{table*}

\subsection{Thresholds \label{sec:thresh}}
The output of the search algorithm outlined in Sec.~\ref{sec:alg} is a $\ml$ value corresponding to the most likely path through each sub-band for each orbital template $(P, a_0, \tasc)$. We flag a template for further follow-up if $\ml$ exceeds a threshold, $\lth$, given an acceptable probability of false alarm. To determine $\lth$ we need to know how often pure noise yields $\ml> \lth$. The distribution of $\ml$ in noise-only data is unknown analytically, but depends on $P$, $a_0$, and the frequency, so Monte-Carlo simulations are used to determine $\lth$ in each sub-band for each target.

We estimate the distribution of $\ml$ in noise via two methods: \begin{enumerate*}[label=\roman*)] \item using realizations of synthetic Gaussian noise generated using the \texttt{lalapps\_Makefakedata\_v5} program in the LIGO Scientific Collaboration Algorithm Library (LALSuite) \cite{LAL2018}, and \item searching O3 data in off-target locations to simulate different realizations of true detector noise.\end{enumerate*} As in Refs.~\cite{o2vitsco, Middleton2020} we generate realizations for each target and sub-band, and apply the search algorithm described in Sec.~\ref{sec:alg} to each realization to recover samples from the noise-only distribution of $\ml$. Details on how we use these samples to find $\lth$ for each sub-band are given in Appendix \ref{app:thresh}. Unless otherwise noted, $\lth$ refers to the lower of the two thresholds derived from the methods listed above to minimize false dismissals.

To define $\lth$ we must also account for a ``trials factor'' due to the number of templates searched in each sub-band. We assume that in noise-only data the spacing between templates is sufficiently large such that each template returns a statistically independent $\ml$. We can therefore relate the false alarm probability for a search of a sub-band with $N_{\rm tot}$ templates, $\alpha_{N_{\rm tot}}$, to the probability of a false alarm for a single template, $\alpha$, viz.~
\begin{equation}
\alpha_{N_{\rm tot}} = 1 - (1 - \alpha)^{N_{\rm tot}}\,. \label{eq:alphan}
\end{equation}
Previous comparable searches have set $\alpha_{N_{\rm tot}}$ between 0.01 and 0.3 \cite{o1crosscorSco, o1vitsco, o2vitsco, Middleton2020}. In this search, we fix $\alpha_{N_{\rm tot}} = 0.3$, i.e.~set the acceptable probability of false alarm at 30\% per sub-band. As we search a total of $20 \times 3 = 60$ sub-bands, we expect $\sim 18$ candidates above $\lth$ due to noise alone (i.e.~false alarms), a reasonable number on which to perform more exhaustive follow-up. Looking ahead to the results in Sec.~\ref{sec:o3results} we recover 4611 candidates above $\lth$. While this number is much higher than the $\sim 18$ false alarms expected, almost all of these candidates are non-Gaussian noise artifacts in one (or both) of the detectors. All but 16 of the 4611 candidates are eliminated by the vetoes outlined in Sec.~\ref{sec:vetoes}. We reiterate that $\lth$ in each sub-band is the lower of the two thresholds described in Appendix \ref{app:thresh}, lowering conservatively the probability of false dismissal. 

\subsection{Computing resources \label{sec:computing}}
A mix of central processing unit (CPU) and graphical processing unit (GPU) resources are used. The GPU implementation of the $\mj$-statistic is identical to that used in Refs.~\cite{o2vitsco, Middleton2020}. The entire search across all targets and sub-bands takes $\sim30$ CPU-hours and $\sim40$ GPU-hours when using compute nodes equipped with Xeon Gold 6140 CPUs and NVIDIA P100 12GB PCIe GPUs. Producing $\lth$ for each sub-band, as described in Sec.~\ref{sec:thresh}, takes an additional $\sim 5\times10^2$ CPU-hours and $\sim 4\times10^3$ GPU-hours to perform the search on different noise realizations. The additional follow-up in Appendix \ref{app:followup} requires an additional $\sim 10^3$ CPU-hours and $\sim 10^2$ GPU-hours.

\section{O3 data \label{sec:o3data}}
We use the full dataset from O3, spanning from April 1, 2019, 15:00 UTC to March 27, 2020, 17:00 UTC, from the LIGO Livingston and Hanford observatories. We do not use any data from the Virgo interferometer in this analysis, due to its lower sensitivity compared to the two LIGO observatories in the frequency sub-bands over which we search \cite{o23DetChar}. The data products ingested by the search algorithm described in Sec.~\ref{sec:alg} are short Fourier transforms (SFTs) lasting $1800\,$s. Times when the detectors were offline, poorly calibrated, or were impacted by egregious noise, are excluded from analysis by using ``Category 1'' vetoes as defined in section 5.2 of Ref.~\cite{o23DetChar}. The SFTs are generated from the ``C01 calibrated self-gated'' dataset, which is the calibrated strain data with loud transient glitches removed \cite{o3gating}. Transient glitches otherwise impact the noise floor, as described in section 6.1 of Ref.~\cite{o23DetChar}. The median systematic error of the strain magnitude across O3 is $<2\%$ \cite{Sun2020, Sun2021}.

The coherence time $\tdr=10\,$d splits the data into $N_T=36$ segments. However, due to the month-long commissioning break between O3a and O3b there are two segments without any SFTs. These two segments, starting at October 8, 2019, 15:00 UTC and October 15, 2019, 15:00 UTC, are replaced with a uniform log-likelihood for all frequency bins, which allows the HMM to effectively skip over them while still allowing spin wandering. When generating synthetic data in Secs.~\ref{sec:thresh} and \ref{sec:ul} the same two data segments are also replaced with uniform log-likelihoods to emulate the real search.

\section{Vetoes \label{sec:vetoes}} 
When a candidate is returned with $\ml > \lth$ we must decide whether there are reasonable grounds to veto the candidate as non-astrophysical. We use three of the vetoes from Ref.~\cite{Middleton2020}: the known line veto, detailed in Sec.~\ref{sec:line_veto}, the single interferometer veto, detailed in Sec.~\ref{sec:sinfo_veto}, and the off-target veto, detailed in Sec.~\ref{sec:ot_veto}. The false dismissal rate of these vetoes is less than $5\%$ (see detailed safety investigations in section IVB of Ref. \cite{o1vitsco} and section IVB of Ref. \cite{o2vitsco}).

\subsection{Known line veto \label{sec:line_veto}}
As part of the detector characterization process many harmonic features are identified as instrumental ``known lines'' \cite{o12lines, o23DetChar}. However, the exact source of these harmonic features is sometimes unidentified, and their impact cannot always be mitigated through isolating hardware components or post-processing the data \cite{o12lines, o23DetChar}. We use the vetted known lines list in Ref.~\cite{o3lineslist}.

Any candidate close to a known line at frequency $f_{\rm line}$ is vetoed. Precisely, if for any time $0 \leq t \leq T_{\rm obs}$ the candidate's frequency path $f(t)$ satisfies 
\begin{equation}
{|f(t) - f_{\rm line}| < 2\pi a_0 f_{\rm line} / P \,,}
\end{equation}
then the candidate is vetoed\footnote{One might consider an additional Doppler broadening factor of $2\pi a_\oplus/1\,$yr, where $a_\oplus$ is the mean Earth-Sun distance, as stationary lines in the detector frame get Doppler shifted when transforming the data to the frame of reference of the source. We opt not to apply this factor for simplicity in this search, as the exact pattern of Doppler modulation depends strongly on the sky location of the target. Looking ahead to the results in Sec.~\ref{sec:o3results}, we note that none of the 16 surviving candidates is within $2 \pi f a_\oplus/1\,$yr of any known line.}. 

\subsection{Single interferometer veto \label{sec:sinfo_veto}}
An instrumental artifact is unlikely to be coincident in both detectors, so the candidate's $\ml$ should be dominated by only one of the detectors if the signal is non-astrophysical. On the other hand, an astrophysical signal may need data from both detectors to be detected, or if it is particularly strong may be seen in both detectors individually. 

We label the original log-likelihood as $\ml_\cup$, and we also calculate the two single interferometer log-likelihoods $\ml_a$ and $\ml_b$ (where the higher $\ml$ is labeled with $b$ for definiteness). There are four possible outcomes for this veto:
\begin{enumerate}
\item If the $\ml$ value in one detector is sub-threshold, while the other is above the two-detector $\ml$ value, i.e.~one has $\ml_a < \lth\ \textrm{and}\ \ml_b > \ml_\cup$ and $f_b(t)$, the frequency path associated with $\ml_b$, is close to the frequency path of the candidate when using data from both detectors, $f_\cup(t)$, i.e.
\begin{equation}
|f_\cup(t) - f_b(t)| < 2\pi a_0 f_\cup / P\,, \label{eq:vetoclose}
\end{equation}
then the candidate is likely to be a noise artifact in detector $b$, and is vetoed.
\item If one has $\ml_a < \lth\ \textrm{and}\ \ml_b > \ml_\cup$, but Eq.~\eqref{eq:vetoclose} does not hold then the candidate signal cannot be vetoed, as the single-interferometer searches did not find the same candidate. This could indicate that the candidate is a weak astrophysical signal that needs data from both detectors to be detectable.
\item If one has $\ml_a > \lth\ \textrm{and}\ \ml_b > \lth$, the candidate could represent a strong astrophysical signal that is visible in data from both detectors independently, or it could represent a common noise source. Candidates in this category cannot be vetoed.
\item If one has $\ml_a < \lth\ \textrm{and}\ \ml_b < \ml_\cup$, data from both detectors is needed for the candidate to be above threshold, possible indicating a weak astrophysical signal. Candidates in this category cannot be vetoed.
\end{enumerate}

\subsection{Off-target veto \label{sec:ot_veto}}
The third veto we apply to a candidate is to search an off-target sky position with the same orbital template. If the off-target search returns $\ml > \lth$ then the candidate is likely instrumental rather than astrophysical. For this veto, off-target corresponds to shifting the target sky position $+40\,$m in RA and $+10\degr$ in Dec.

\section{O3 search results \label{sec:o3results}} 
\begin{figure*}
	\centering
	\includegraphics[width=0.8\linewidth]{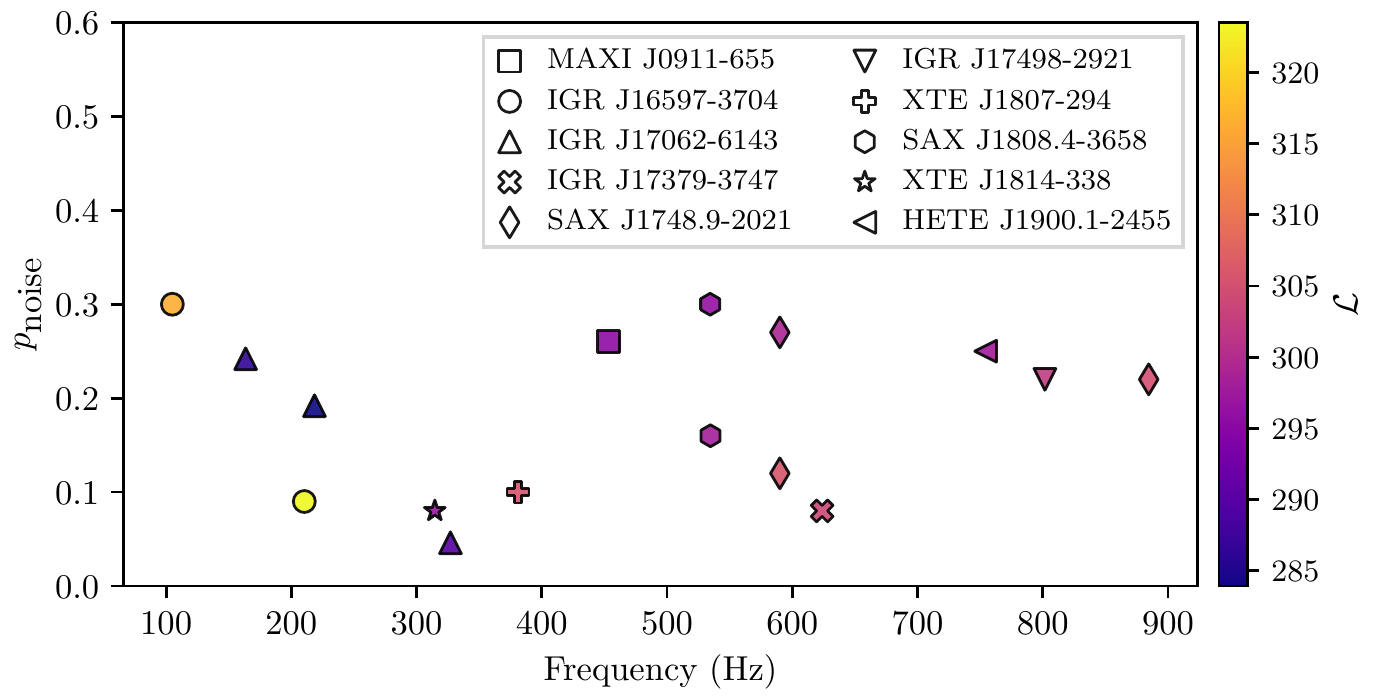}
	\caption{Summary of search results across all targets and sub-bands with $\ml > \lth$. The different symbols correspond to candidates from different targets. The ordinate shows $\pn$ for each candidate, the probability that a search of that candidate's sub-band in pure noise would return at least one candidate at least as loud as the one seen. The color of each candidate indicates $\mathcal{L}$ (see color bar at right). Candidates that are eliminated by the vetoes outlined in Sec.~\ref{sec:vetoes} are not shown for clarity. Details on the search results are in Sec.~\ref{sec:o3results} and Appendix \ref{app:fullresults}.}
	\label{fig:summary}
\end{figure*}

The results of the search of all 20 targets are summarized in Fig.~\ref{fig:summary}, with $\alpha_{N_{\rm tot}}=0.3$, i.e.~a nominal probability of false alarm per sub-band of 30\%. Each symbol indicates, for all templates with $\ml > \lth$, the terminating frequency bin and $\pn$, the probability that a search of that candidate's sub-band in pure noise would return at least one candidate at least as loud as the one seen. Equation \eqref{eq:pn} in Appendix \ref{app:pnoise} defines $\pn$ explicitly. Each candidate is colored according to $\mathcal{L}$. We note that high $\mathcal{L}$ does not always correspond to low $\pn$ due to the differing ``trials factors'' in each sub-band, as accounted for when calculating $\lth$ via Eq.~\eqref{eq:alphan}. A low value of $\pn$ corresponds to a higher probability that the candidate is a true astrophysical signal. Targets not listed in the legend return zero candidates above threshold. We do not display in Fig.~\ref{fig:summary} candidates that are eliminated by any of the vetoes described in Sec.~\ref{sec:vetoes} for clarity. 

In total, across all targets and sub-bands, there are 4611 candidates with $\ml > \lth$, before the vetoes are applied. All but 100 are eliminated by veto A (known line veto). A further 84 candidates are eliminated by veto B (single interferometer veto). None of the remaining candidates are eliminated by veto C (off-target veto), leaving 16 candidates passing all of the vetoes outlined in Sec.~\ref{sec:vetoes}. None of the surviving candidates from the O3 search coincide in their orbital template and terminating frequency bin with the seven above- or sub-threshold candidates from the O2 search (c.f. Table VI of Ref.~\cite{Middleton2020}). If we set $\alpha_{N_{\rm tot}}=0.01$, i.e.~set the probability of false alarm per sub-band to 1\%, the search does not return any candidates with $\ml>\lth$ for any target or sub-band, after vetoes are applied.

In Secs.~\ref{sec:igrh}--\ref{sec:hete} we summarize the search results for each of the 20 targets. To guide the reader, and not clutter the main body of the paper, the full search results for one target, \igrh, are shown in Fig.~\ref{fig:igrh}, while the full search results for the other 19 targets are shown in Figs.~\ref{fig:igra}--\ref{fig:hete} in Appendix \ref{app:fullresults}. The orbital template, terminating frequency bin, $\ml$, and $\pn$ for all 16 candidates with $\ml > \lth$ are collated in Table \ref{tab:outliers} in Appendix \ref{app:fullresults}. We present further follow-up of the 16 candidates in Appendix \ref{app:followup}. We find no convincing evidence that any are a true astrophysical signal.

\subsection{\igrh  \label{sec:igrh} }
\begin{figure*}
	\centering
	\includegraphics[width=0.9\linewidth]{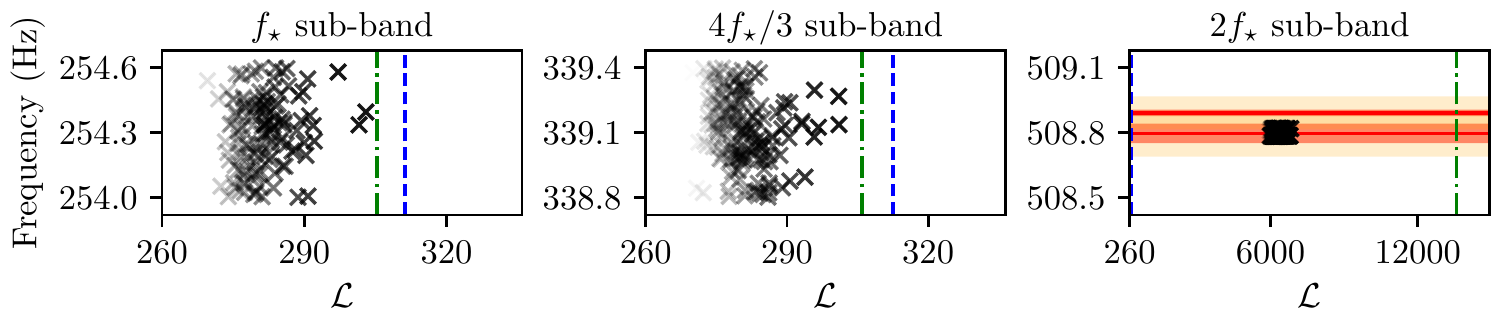}
	\caption{Search results for \igrh. Black crosses indicate the terminating frequency and $\ml$ for the most likely path through the sub-band for each binary template. The vertical blue dashed (green dot-dashed) lines correspond to the threshold set via Gaussian (off-target) noise realizations, $\lthg$ ($\lthot$), in each sub-band. Solid red lines in the right panel indicate the peak frequency of known instrumental lines in the Hanford detector; the orange band indicates the width of the line in the detector frame and the yellow band indicates the increased effective width due to Doppler broadening, as described in Sec.~\ref{sec:line_veto}. Multiple overlapping orange bands creates the red bands. The sub-band around $508.8\,$Hz is especially noisy due to test mass suspension violin mode resonances \cite{o23DetChar}. The transparency of crosses in sub-bands with many templates, is adjusted relative to the maximum $\ml$ in that sub-band for clarity.}
	\label{fig:igrh}	
\end{figure*}
The search results for \igrh\ are presented in Fig.~\ref{fig:igrh}. Each marker in Fig.~\ref{fig:igrh} shows the terminating frequency and associated $\ml$ of the most likely path through the sub-band for a given template, i.e.~choice of $P$ and $\tasc$. The vertical blue dashed (green dot-dashed) lines correspond to the threshold set via Gaussian (off-target) noise realizations, $\lthg$ ($\lthot$), in each sub-band, with $\alpha_{N_{\rm tot}} = 0.3$. See Appendix \ref{app:thresh} for details on how we set thresholds in each sub-band. The horizontal red lines indicate known instrumental lines in the detector with bandwidth indicated by the shading. There are zero above-threshold candidates in the $f_\star$ and $4f_\star / 3$ sub-bands. There are 435 above-threshold candidates in the $2f_\star$ sub-band, which are all coincident with known noise lines in both the Livingston and Hanford detectors, and are therefore eliminated by veto A. The sub-band around $508\,$Hz is especially noisy due to violin mode resonances \cite{o23DetChar}.

\subsection{\igra  \label{sec:igra} }
The search results for \igra\ are shown in Fig.~\ref{fig:igra}, which is laid out identically to Fig.~\ref{fig:igrh}. There are zero above-threshold candidates in the $4f_\star / 3$ and $2f_\star$ sub-bands. There are three above-threshold candidates in the $f_\star$ sub-band, however all three of these candidates are coincident with known noise lines in the Hanford detector, and are therefore eliminated with veto A.

\subsection{\maxi  \label{sec:maxi} }
The search results for \maxi\ are shown in Fig.~\ref{fig:maxi}, which is laid out identically to Fig.~\ref{fig:igrh}. There are zero above-threshold candidates in the $f_\star$ and $2f_\star$ sub-bands. There is one above-threshold candidate in the $4f_\star / 3$ sub-band which survives all of the vetoes and has $\pn=0.26$. Additional follow-up, presented in Appendix \ref{app:followup}, does not provide any evidence that this candidate is a true astrophysical signal.

\subsection{\xtea  \label{sec:xtea} }
The search results for \xtea\ are shown in Fig.~\ref{fig:xtea}, which is laid out identically to Fig.~\ref{fig:igrh}. There are zero above-threshold candidates across all three sub-bands.

\subsection{\igrb  \label{sec:igrb} }
The search results for \igrb\ are shown in Fig.~\ref{fig:igrb}, which is laid out identically to Fig.~\ref{fig:igrh}. Each sub-band for this target is contaminated with known noise lines. There are 84 above-threshold candidates in the $4f_\star / 3$ sub-band, however they are all eliminated by veto B. One above-threshold candidate is returned in each of the $f_\star$ and $2f_\star$ sub-bands. Both of these candidates survive all of the vetoes, and have $\pn=0.30$ and $\pn=0.09$ respectively. Further follow-up, including the frequency path and cumulative log-likelihood for the latter candidate, is presented in Appendix \ref{app:followup}. This follow-up does not provide any evidence that either candidate is a true astrophysical signal.

\subsection{\igrc  \label{sec:igrc} }
The search results for \igrc\ are shown in Fig.~\ref{fig:igrc}, which is laid out identically to Fig.~\ref{fig:igrh}. Given the long-term timing presented in Ref.~\cite{Bult2021} there is only one template needed in each of the three sub-bands for this target. The template returns $\ml > \lth$ in all three of the $f_\star$, $4f_\star / 3$, and $2f_\star$ sub-bands. All of these candidates survive all of the vetoes, and have $\pn=0.24$, $\pn=0.19$, and $\pn=0.05$ respectively. Further follow-up, including the frequency path and cumulative log-likelihood for the candidate with $\pn=0.05$, is presented in Appendix \ref{app:followup}. This follow-up does not provide any evidence that any of the three candidates are a true astrophysical signal.

\subsection{\igrd  \label{sec:igrd} }
The search results for \igrd\ are shown in Fig.~\ref{fig:igrd}, which is laid out identically to Fig.~\ref{fig:igrh}. There are zero above-threshold candidates in the $f_\star$ and $2f_\star$ sub-bands. There is one above-threshold candidate in the $4f_\star/3$ sub-band which survives all of the vetoes and has $\pn=0.08$. Further follow-up, including the frequency path and cumulative log-likelihood, for this candidate is presented in Appendix \ref{app:followup}. This follow-up does not provide any evidence that the candidate is a true astrophysical signal.

\subsection{\saxb   \label{sec:saxb} }
The search results for \saxb\ are shown in Fig.~\ref{fig:saxb}, which is laid out identically to Fig.~\ref{fig:igrh}. There are zero above-threshold candidates in the $f_\star$ sub-band. There are two above-threshold candidates in the $4 f_\star / 3$ sub-band which survive all of the vetoes and have $\pn=0.12$ and $\pn=0.27$. There is one above-threshold candidate in the $2 f_\star$ sub-band which survives all of the vetoes and has $\pn=0.22$. Additional follow-up, presented in Appendix \ref{app:followup}, does not provide any evidence that any of the three candidates are a true astrophysical signal.

\subsection{\ngc   \label{sec:ngc} }
The search results for \ngc\ are shown in Fig.~\ref{fig:ngc}, which is laid out identically to Fig.~\ref{fig:igrh}. There are zero above-threshold candidates across all three sub-bands.

\subsection{\igri  \label{sec:igri} }
The search results for \igri\ are shown in Fig.~\ref{fig:igri}, which is laid out identically to Fig.~\ref{fig:igrh}. There are zero above-threshold candidates in the $f_\star$ and $2f_\star$ sub-bands. All 4050 candidates in the $4f_\star/3$ sub-band are above threshold, however all of them are coincident with a known noise line in the Hanford detector, and are therefore eliminated with veto A. The sub-band around $501.7\,$Hz is especially noisy due to violin mode resonances \cite{o23DetChar}.

\subsection{\swiftb \label{sec:swiftb} }
The search results for \swiftb\ are shown in Fig.~\ref{fig:swiftb}, which is laid out identically to Fig.~\ref{fig:igrh}. There are zero above-threshold candidates in the $f_\star$ and $4f_\star / 3$ sub-bands. There is one above threshold candidate in the $2 f_\star$ sub-band. However it is coincident with a known noise line in the Hanford detector, and is therefore eliminated by veto A.

\subsection{\igre  \label{sec:igre} }
The search results for \igre\ are shown in Fig.~\ref{fig:igre}, which is laid out identically to Fig.~\ref{fig:igrh}. There are zero above-threshold candidates in the $f_\star$, and $4f_\star / 3$ sub-bands. There is one above-threshold candidate in the $2f_\star$ sub-band which survives all of the vetoes and has $\pn=0.22$. Additional follow-up, presented in Appendix \ref{app:followup}, does not provide any evidence that this candidate is a true astrophysical signal.

\subsection{\igrf  \label{sec:igrf} }
The search results for \igrf\ are shown in Fig.~\ref{fig:igrf}, which is laid out identically to Fig.~\ref{fig:igrh}. There are zero above-threshold candidates across all three sub-bands.

\subsection{\xteb  \label{sec:xteb} }
The search results for \xteb\ are shown in Fig.~\ref{fig:xteb}, which is laid out identically to Fig.~\ref{fig:igrh}. There are zero above-threshold candidates across all three sub-bands.

\subsection{\swift \label{sec:swift} }
The search results for \swift\ are shown in Fig.~\ref{fig:swift}, which is laid out identically to Fig.~\ref{fig:igrh}. There are zero above-threshold candidates across all three sub-bands.

\subsection{\igrg  \label{sec:igrg} }
The search results for \igrg\ are shown in Fig.~\ref{fig:igrg}, which is laid out identically to Fig.~\ref{fig:igrh}. There are zero above-threshold candidates across all three sub-bands.

\subsection{\xtec  \label{sec:xtec} }
The search results for \xtec\ are shown in Fig.~\ref{fig:xtec}, which is laid out identically to Fig.~\ref{fig:igrh}. There are zero above-threshold candidates in the $f_\star$ and $4f_\star / 3$ sub-bands. There is one above-threshold candidate in the $2f_\star$ sub-band which survives all of the vetoes and has $\pn=0.10$. Further follow-up, including the frequency path and cumulative log-likelihood, for this candidate is presented in Appendix \ref{app:followup}. This follow-up does not provide any evidence that the candidate is a true astrophysical signal.

\subsection{\sax   \label{sec:sax} }
The search results for \sax\ are shown in Fig.~\ref{fig:sax}, which is laid out identically to Fig.~\ref{fig:igrh}. There are zero above-threshold candidates in the $f_\star$ and $2f_\star$ sub-bands. There are two above-threshold candidates in the $4f_\star / 3$ sub-band which survive all of the vetoes and have $\pn=0.16$ and $\pn=0.30$. Additional follow-up, presented in Appendix \ref{app:followup}, does not provide any evidence that either candidate is a true astrophysical signal.

\sax\ was observed in outburst in August 2019, during O3a \cite{Bult2020, Goodwin2020}. This allows us to perform an additional target-of-opportunity search during only its active phase. If the target only emits continuous gravitational waves during outburst, searching a shorter duration of data increases the probability of detection by increasing the signal-to-noise ratio. The details and results of this target-of-opportunity search are in Sec.~\ref{sec:shortsax}. In summary, after searching with three separate coherence times of $\tdr = 1\,$d, $\tdr=8\,$d, and $\tdr=24\,$d, only one candidate is above threshold and survives all of the vetoes. The candidate is found using $\tdr=24\,$d in the $f_\star$ sub-band, and has $\pn = 0.02$. Additional follow-up does not reveal any informative features that would distinguish between an astrophysical signal and noise. It does not coincide with either of the two candidates in the $4f_\star / 3$ sub-band found in the semi-coherent search using the full O3 data set. 

\subsection{\xted  \label{sec:xted} }
The search results for \xted\ are shown in Fig.~\ref{fig:xted}, which is laid out identically to Fig.~\ref{fig:igrh}. There are zero above-threshold candidates in the $4f_\star / 3$ and $2f_\star$ sub-bands. There is one above-threshold candidate in the $f_\star$ sub-band which survives all of the vetoes and has $\pn=0.08$.  Further follow-up, including the frequency path and cumulative log-likelihood, for this candidate is presented in Appendix \ref{app:followup}. This follow-up does not provide any evidence that the candidate is a true astrophysical signal.

\subsection{\hete  \label{sec:hete} }
The search results for \hete\ are shown in Fig.~\ref{fig:hete}, which is laid out identically to Fig.~\ref{fig:igrh}. There are zero above-threshold candidates in the $f_\star$ sub-band. All 22 templates in the $4f_\star / 3$ sub-band return candidates above $\lth$, however these candidates are all coincident with known noise lines in the Hanford detector, and are summarily eliminated with veto A. The sub-band around $503\,$Hz is especially noisy due to violin mode resonances \cite{o23DetChar}. There is one above-threshold candidate in the $2f_\star$ sub-band which survives all of the vetoes and has $\pn=0.25$. Additional follow-up, presented in Appendix \ref{app:followup}, does not provide any evidence that this candidate is a true astrophysical signal.

\section{Target-of-opportunity search: \sax\ in outburst \label{sec:shortsax}}
On August 7 2019 \sax\ went into outburst \cite{Bult2019}. The Neutron star Interior Composition Explorer (NICER) team undertook a high-cadence monitoring campaign, and performed a timing analysis of the pulsations \cite{Bult2020}. The outburst lasted for roughly 24 days, with enhanced X-ray flux observed between August 7 2019 and August 31 2019 (see Fig.~1 of Ref.~\cite{Bult2020}). We note that the \emph{Swift} X-ray Telescope observed increased X-ray activity from August 6 2019, and observations in the optical $i'$-band with the Las Cumbres Observatory network detected an increased flux from July 25 2019 \cite{Goodwin2020}.

Outburst events are attributed to in-falling plasma that is channeled by the magnetosphere onto a localized region on the neutron star surface, creating a hot spot that rotates with the star \cite{Romanova2004}. As the observed X-ray flux is assumed to be linearly proportional to the mass accretion rate, an outburst could result in a larger mountain on the neutron star surface (or excite $r$-modes in the interior), compared to when the AMXP is in quiescence \cite{Haskell2015GW, Haskell2017a}. 

If continuous gravitational waves are only emitted from \sax\ when it is in outburst, searching all of the O3 data decreases the signal-to-noise ratio, as compared to only searching data from the outburst. To protect against this possibility, we do an additional search for continuous gravitational waves from \sax\ using data from both LIGO observatories between 1249171218 GPS time (August 7 2019) and 1251244818 GPS time (August 31 2019), rather than data from the entirety of O3, as in Sec.~\ref{sec:sax}. 

\subsection{Search parameters}
The search algorithm is laid out in Sec.~\ref{sec:alg}. We run the search using three different coherence times, setting $\tdr = 1\,$d, $\tdr = 8\,$d, and $\tdr = 24\,$d. We search three sub-bands centered on $\{1, 4/3, 2\}f_\star$, for each $\tdr$. The width of the sub-band depends on $\tdr$. It is $\sim 0.76\,$Hz for the searches with $\tdr=1\,$d and $8\,$d, and is $\sim 1.01\,$Hz for the search with $\tdr=24\,$d. Given the precise timing achieved during the outburst in 2019 \cite{Bult2020}, and the shorter search duration, only one $\{P,\ \tasc,\ a_0\}$ template is required for each sub-band, according to Eqs.~\eqref{eq:np}--\eqref{eq:ntasc}. Due to the different values of $\tdr$, shorter total duration, and different number of templates, we re-calculate $\lth$ for each sub-band and value of $\tdr$, using the procedure outlined in Sec.~\ref{sec:thresh} and Appendix \ref{app:thresh}. As in the full O3 search, we set the probability of false alarm in each sub-band at $\alpha_{N_{\rm tot}} = 0.3$. For all candidates that have $\ml > \lth$ we apply the three vetoes described in \ref{sec:vetoes}.

\subsection{Search results}
For $\tdr = 1\,$d, the search in the $f_\star$ sub-band returns one candidate above $\lth$. The candidate survives both veto A (known line) and veto B (single interferometer), but fails veto C (off-target). The searches in the $4/3f_\star$ and $2f_\star$ sub-bands do not return any candidates above $\lth$.

For $\tdr = 8\,$d, there are no candidates above $\lth$ in any of the three sub-bands.

For $\tdr = 24\,$d, the searches in the $4 f_\star /3$ and $2f_\star$ sub-bands do not return any candidates above $\lth$. The search in the $f_\star$ sub-band does return one candidate above $\lth$. This candidate survives all of the vetoes outlined in Sec.~\ref{sec:vetoes}. We remind the reader that with $\alpha_{N_{\rm tot}} = 0.3$ and nine sub-bands searched (three for each of the three choices of $\tdr$), we should expect $\sim 3$ candidates above threshold purely due to noise. The probability that we would see a value of $\ml$ at least this large if this sub-band is pure noise, $\pn$, is 0.02. The template and frequency of the candidate are not coincident with any candidate from the full O3 search (see Table \ref{tab:outliers}) or the sub-threshold candidate found in the search of this sub-band in O2 data \cite{Middleton2020}. By setting $\tdr = T_{\rm obs} = 24\,$d we perform a fully coherent search across this time period, with a frequency bin size of $\Delta f = 2.4\times10^{-7}\,$Hz. We describe in Appendix \ref{app:shortsax_followup} further follow-up of this candidate. In summary, we find no significant evidence that it is an astrophysical signal rather than a noise fluctuation.

\section{Frequentist upper limits \label{sec:ul}}
If we assume that the remaining candidates reported in Sec.~\ref{sec:o3results} and Appendix \ref{app:fullresults} are false alarms, we can place an upper limit on the wave strain that is detectable at a confidence level of 95\%, $\hul$, in a sub-band. The value of $\hul$ is a function of our algorithm, the detector configuration during O3, and our assumptions about the signal model. We describe the method used to estimate $\hul$ in Sec.~\ref{sec:ul_method}, present the upper limits in each sub-band in Sec.~\ref{sec:ul_results}, and compare the results to indirect methods that calculate the expected strain in the $2f_\star$ sub-band in Sec.~\ref{sec:ul_comp}. The astrophysical implications are discussed in Sec.~\ref{sec:ul_impl}.

\subsection{Upper limit procedure in a sub-band \label{sec:ul_method}}
We set empirical frequentist upper limits in each sub-band using a sequence of injections into O3 SFTs. For each sub-band we inject $N_{\rm trials}=100$ simulated binary signals at 12--15 fixed values of $h_0$ using \texttt{lalapps\_Makefakedata\_v5} \cite{LAL2018}. For each of the $N_{\rm trials}$ injections at a fixed $h_0$ we select a constant injection frequency, $f_{\rm inj}$, uniformly from the sub-band. While the injected signal has zero spin-wandering, we still use $\tdr=10\,$d in the search algorithm outlined in Sec.~\ref{sec:alg} to mimic the real search. The injected period, $P_{\rm inj}$, and time of ascension, $T_{\rm asc,\,inj}$ are chosen uniformly from the ranges $[P - 3\sigma_P, P + 3\sigma_P]$ and $[\tasc - 3\sigma_{\tasc}, \tasc + 3\sigma_{\tasc}]$ respectively. We keep $a_0$ fixed at the precisely known value for each target. The polarization, $\psi$, is chosen uniformly from the range $[0, 2\pi]$. The cosine of the projected inclination angle of the neutron star spin axis with our line of sight, $\cos\iota$, is chosen uniformly from the range $[-1, 1]$\footnote{While the inclination angle of the binary with respect to our line of sight is restricted via EM observations for some of our targets, we opt to marginalize over $\cos\iota$ as the neutron star spin axis may not necessarily align with the orbital axis of the binary. It is possible to scale our results via equation (19) of Ref.~\cite{Messenger2015a}, if one wishes to fix $\cos\iota$.}. We then search for the injected signal with the template in this sub-band's template grid that is nearest to $\{P_{\rm inj}, T_{\rm asc,\,inj}\}$. We re-calculate $\lth$ such that the probability of false alarm in each sub-band is $\alpha_{N_{\rm tot}} = 0.01$. This allows us to set conservative upper limits, even in sub-bands where we have marginal candidates above a threshold corresponding to a probability of false alarm of 30\% per sub-band. By recording the fraction of injected signals we recover at each $h_0$ with $\ml > \lth$ we estimate the efficiency, $\varepsilon$, as a function of $h_0$. We then perform a logistic regression \cite{Gelman2013} to obtain a sigmoid fit to $\varepsilon(h_0)$, and solve 
\begin{equation}
\varepsilon(\hul) = 0.95\ , \label{eq:effic_95}
\end{equation}
to find an estimate of $\hul$ in the given sub-band. 

One might reasonably ask, how precise is this estimate of $\hul$? The main factors impacting the precision are: \begin{enumerate*}[label=(\roman*)]
\item the precision of the most likely parameters of the sigmoid, as estimated via logistic regression, when solving Eq.~\eqref{eq:effic_95} for $\hul$, given the $N_{\rm trials}$ injections done at 12--15 values of $h_0$; and
\item the assumption that the strain data (and hence the SFTs) are perfectly calibrated.
\end{enumerate*}
We investigate the impact of (i) by drawing alternative sigmoid fits of $\varepsilon(h_0)$ using the covariance matrix of the parameters returned by the logistic regression. We find that inverting these alternative fits through Eq.~\eqref{eq:effic_95} results in a value of $\hul$ that varies by less than $5\%$ from the value calculated via the most likely parameters (at the 95\% confidence level). The impact of (ii) is trickier to quantify. As described in Refs.~\cite{Sun2020, Sun2021} the median systematic error in the magnitude of the strain is less than $2\%$ in the 20--2000\,Hz frequency band across O3a. The statistical uncertainty around the measurement of calibration bias means that in the worst case the true magnitude of the calibration bias may be as large as $7\%$. However, the calibration bias at a given frequency is not correlated between the detectors (see Figures 16 and 17 in Ref.~\cite{Sun2020}), and so the impact on a continuous gravitational wave search that combines data from both detectors is likely to be less than $7\%$. 

In light of the above considerations we quote $\hul$ to a precision of two significant figures, but we emphasize that estimating $\hul$ involves many (potentially compounding) uncertainties. Subsequent conclusions about the physical system that are drawn from estimates of $\hul$ cannot be more precise than the estimate of $\hul$ itself.

\subsection{Upper limits \label{sec:ul_results}}
The estimates of $\hul$ for each target and sub-band are listed in Table \ref{tab:ul}. Dashes correspond to sub-bands that are highly contaminated with noise lines, which preclude the procedure described in Sec.~\ref{sec:ul_method}, as one always finds $\ml > \lth$, regardless of $h_0$. The most sensitive sub-bands are for IGR J17062$-$6143 with $\hul = 4.7 \times 10^{-26}$ in both the $4f_\star/3$ and $2f_\star$ sub-bands (centered around 218.2\,Hz and 327.6\,Hz respectively). These sub-bands lie in the most sensitive band of the detector, and the binary elements are known to high precision \cite{Bult2021}, so only one template is needed in each sub-band, corresponding to a relatively lower $\lth$ at fixed probability of false alarm.

No estimates of $\hul$ were established in Ref.~\cite{Middleton2020} for the five targets therein. The search of XTE J1751$-$305 in S6 data estimated $\hul \approx 3.3\times10^{-24}$, $4.7\times10^{-24}$, and $7.8\times10^{-24}$ in three sub-bands corresponding to $f_\star$, an $r$-mode frequency, and $2f_\star$ respectively \cite{s6twoSpectScoXTE}. Our estimates of $\hul$ for XTE J1751$-$305 improve these results by two orders of magnitude, because the detector is more sensitive, and $\tdr$ is longer.

\begin{table}[t]
\caption{Upper limits on the detectable gravitational wave strain at a 95\% confidence level, $\hul$, in each of the sub-bands for each target. See Sec.~\ref{sec:ul_method} for details on how they are estimated, and the precision to which they are known. Upper limits are not estimated in sub-bands marked with a ``$-$'' as these sub-bands are highly contaminated with known noise lines. \label{tab:ul}}
\begin{ruledtabular}
\begin{tabular}{l d{5} d{5} d{5}}
				& \multicolumn{3}{c}{\textrm{$\hul$ in each sub-band ($\times 10^{-26}$)}} \\
\textrm{Target}	& \textrm{$f_{\rm \star}$} & \textrm{$4f_{\rm \star}/3$} & \textrm{$2f_{\rm \star}$}  \\
\midrule
IGR J00291$+$5934      & -  	    & 7.6     	& 11    \\
MAXI J0911$-$655       & 7.7  	    & 6.4     	& 7.3   \\
XTE J0929$-$314        & 5.1  	    & 5.3       & 6.4   \\
IGR J16597$-$3704      & 7.5  	    & - 		& 5.6   \\
IGR J17062$-$6143      & 8.1  	    & 4.7   	& 4.7   \\
IGR J17379$-$3747      & 8.5  	    & 7.4   	& 10    \\
SAX J1748.9$-$2021     & 9.2  	    & 7.7   	& 10  \\
NGC 6440 X$-$2         & 6.2  	    & 7.2     	& 5.8   \\
IGR J17494$-$3030	   & 8.3  	    & - 		& 9.0   \\
Swift J1749.4$-$2807   & 11 	    & 17  		& 24    \\
IGR J17498$-$2921      & 7.0  	    & 6.6       & 8.4   \\
IGR J17511$-$3057      & 7.5  	    & 5.5       & 6.6   \\
XTE J1751$-$305        & 10 	    & 8.3       & 9.7   \\
Swift J1756.9$-$2508   & 8.1  	    & 8.8     	& 6.3   \\
IGR J17591$-$2342      & 9.5  	    & 11   	    & 14    \\
XTE J1807$-$294        & 6.1  	    & 5.0       & 5.6   \\
SAX J1808.4$-$3658     & 6.4        & 6.9       & 8.8   \\
XTE J1814$-$338        & 9.4  	    & 6.0       & 6.9   \\
IGR J18245$-$2452      & 9.0  	    & 6.3       & - 	\\
HETE J1900.1$-$2455    & 5.6    	& -	        & 8.4   \\
\end{tabular}
\end{ruledtabular}
\end{table}

\subsection{Comparison to expected strain from AMXPs \label{sec:ul_comp}}

\begin{table*}[t]
\begin{threeparttable}
\caption{Maximum expected strain from each target, as inferred from EM observations. The second column contains the best estimate for the distance to the target. Targets with ``-'' listed as the frequency derivative (third column), $\dot{f}_\star$, do not have a measured value during outburst, and also do not have a long-term (quiescent) $\dot{f}_\star$ measured either. The labels (A) and (Q) indicate that $\dot{f}_\star$ is measured in outburst and quiescence respectively. The scaling equations used to estimate the maximum spin-down strain (fourth column), $h_{0,\,\textrm{sd}}$, and the maximum strain assuming torque-balance (sixth column), $h_{0\,\textrm{torque}}$, are Eqs.~\eqref{eq:h0sd} and \eqref{eq:h0t} respectively. The $h_{0,\,\textrm{sd}}$ value is calculated using the central distance and $\dot{f}_\star$ estimates. The $h_{0\,\textrm{torque}}$ value is calculated using the maximum bolometric X-ray flux measured during outburst (fifth column), $F_{X,\,\textrm{max}}$, which is typically measured to a precision of $\sim10\%$. The X-ray flux of each target in quiescence is not shown, as it is only measured for half of the targets, and is usually $\sim1-2$ orders of magnitude lower than $F_{X,\,\textrm{max}}$. The seventh column contains $\hul$ in the $2f_\star$ sub-band (fourth column of Table \ref{tab:ul}) to facilitate comparisons between $\hul$ and $h_{0\,\textrm{torque}}$ or $h_{0,\,\textrm{sd}}$. \label{tab:exp_ul}}
\begin{ruledtabular}
\begin{tabular}{l r r r r r r l}
	   & \textrm{Distance} &	 & \textrm{$h_{0,\,\textrm{sd}}$} & \textrm{$F_{X,\,\textrm{max}}$ $(\times 10^{-8}$} & \textrm{$h_{0,\textrm{torque}}$} & \textrm{$\hul$} & \\
Target & \textrm{(kpc)} & \textrm{$\dot{f}_\star$ (Hz\,s$^{-1}$)}  & \textrm{$(\times 10^{-26})$} & \textrm{erg\,s$^{-1}\,$cm$^{-2}$)} & \textrm{$(\times 10^{-26})$} & \textrm{$(\times 10^{-26})$} & \textrm{Refs.} \\
\midrule
IGR J00291$+$5934   	& \textrm{4.2(5)} 		 & $-4.0(1.4)\times 10^{-15}$  (Q)				& 0.05 			& 0.35  	& 0.2   & 11   &\cite{Watts2008, Papitto2011, Sanna2017, DeFalco2017} \\
MAXI J0911$-$655    	& \textrm{9.45(15)} 	 & -	  										& -				& 0.047  	& 0.1   & 7.3  &\cite{Watkins2015, Homan2016, Sanna0911} \\
XTE J0929$-$314     	& \textrm{7.4\tnote{a}}  & $-9.2(4)\times 10^{-14}$  (A)  				& 0.2			& 0.1  		& 0.2   & 6.4  &\cite{Galloway2002, Watts2008, Marino2017} \\
IGR J16597$-$3704   	& \textrm{9.1\tnote{b}}  & -  						   	  				& -				& 0.065  	& 0.2   & 5.6  &\cite{Sanna16597} \\
IGR J17062$-$6143   	& \textrm{7.3(5)} 		 & $+3.77(9)\times 10^{-15}$ (A)  				& 0.04\tnote{f}	& 0.006  	& 0.05  & 4.7  &\cite{Keek2017, Bult2021} \\
IGR J17379$-$3747   	& \textrm{8\tnote{c}} 	 & $-1.2(1.9)\times 10^{-14}$\tnote{e} (A)		& 0.05			& 0.04  	& 0.08  & 10   &\cite{VanDenEijnden2018a, Negoro2018, Sanna17379, Bult17379} \\
SAX J1748.9$-$2021  	& \textrm{8.5\tnote{b}}  & -										  	& -				& 0.077  	& 0.1   & 10   &\cite{Watts2008, Harris2010, Sanna1748, Sharma2020} \\
NGC 6440 X$-$2      	& \textrm{8.5\tnote{b}}  & -  						   			    	& -				& 0.02  	& 0.09  & 5.8  &\cite{Harris2010, Bult2015} \\
IGR J17494$-$3030     	& \textrm{8\tnote{c}} 	 & $-2.1(7)\times 10^{-14} $ (Q)  				& 0.07			& 0.0143  	& 0.05  & 9.0  &\cite{Ng2021} \\
Swift J1749.4$-$2807	& \textrm{6.7(1.3)} 	 & -  						   			    	& -				& 0.0352  	& 0.07  & 24   &\cite{Wijnands2009, Bult2021a, Sanna1749} \\
IGR J17498$-$2921   	& \textrm{7.6(1.1)}  	 & $-6.3(1.9)\times 10^{-14}\ \tnote{e}\ $ (A) 	& 0.1 			& 0.2  		& 0.2   & 8.4  &\cite{Linares2011, Papitto17498, Falanga2012} \\
IGR J17511$-$3057   	& \textrm{3.6(5)} 		 & $+4.8(1.4)\times 10^{-14} $ (A)  			& 0.2\tnote{f}	& 0.2  		& 0.2   & 6.6  &\cite{Riggio17511} \\
XTE J1751$-$305     	& \textrm{6.7\tnote{d}}  & $-5.5(1.2)\times 10^{-15}$  (Q)  			& 0.04			& 0.29  	& 0.2   & 9.7  &\cite{Watts2008, Papitto1751, Riggio2011} \\
						& 						 & $+3.7(1.0)\times 10^{-13}$  (A)  			& 0.2\tnote{f}	& 		  	& 	    &      & \\
Swift J1756.9$-$2508	& \textrm{8\tnote{c}} 	 & $-4.8(6)\times 10^{-16}$  (Q)  				& 0.02			& 0.288  	& 0.3   & 6.3  &\cite{Watts2008, Sanna1756} \\
						& 					 	 & $-4.3(2.1)\times 10^{-11}$\ \tnote{e}\ \ (A) & 5				& 		  	& 	    &      & \\
IGR J17591$-$2342   	& \textrm{7.6(7)} 		 & $-7.1(4)\times 10^{-14} $ (A)  				& 0.1			& 0.0535  	& 0.09  & 14   &\cite{Gusinskaia2020, Kuiper2020, Sanna17591} \\
XTE J1807$-$294     	& \textrm{8\tnote{c}} 	 & $+2.7(1.0)\times 10^{-14}$  (A)  			& 0.08\tnote{f}	& 0.2  		& 0.3   & 8.8  &\cite{Watts2008, Riggio2008, Patruno1807} \\
SAX J1808.4$-$3658  	& \textrm{$3.3^{+0.3}_{-0.2}$} & $-1.01(7)\times 10^{-15}$ (Q)  		& 0.04			& 0.103  	& 0.1   & 5.6  &\cite{Galloway2006, Watts2008, Goodwin2019, Bult2020} \\
						& 				 		 & $-3.02(13)\times 10^{-13}$ (A) & 0.7			& 		  	& 	    & 	   & \\
XTE J1814$-$338     	& \textrm{10.25(1)} 	 & $-6.7(7)\times 10^{-14}$  (A)  				& 0.1			& 0.069  	& 0.1   & 6.9  &\cite{Papitto1814, Watts2008, DAvanzo2009} \\
IGR J18245$-$2452   	& \textrm{5.5\tnote{b}}  & -  						   			    	& -				& 0.0466  	& 0.1   & -    &\cite{Harris2010, Papitto18245, Campana2018} \\
HETE J1900.1$-$2455 	& \textrm{4.5(2)} 		 & $+4.2(1)\times 10^{-13}$  (A)  				& 0.4\tnote{f}	& 0.09  	& 0.1   & 8.4  &\cite{Suzuki2007, Galloway2008, Patruno1900, Watts2008} \\
\end{tabular}
\end{ruledtabular}
\begin{tablenotes}
\item[a]{Estimate assumes conservative mass transfer during accretion. An alternative estimate gives less than $4\,$kpc \cite{Marino2017}.}
\item[b]{Uncertainty not quoted as target located in a globular cluster.}
\item[c]{Unknown, but as the target is in the direction of the galactic centre a fiducial value of 8\,kpc is assumed in the literature.}
\item[d]{Lower limit.}
\item[e]{Estimate of $\dot{f}_\star$ consistent with zero at a $3\sigma$ level.}
\item[f]{Assumes $\dot{f}_\textrm{GW} \approx - \dot{f}_\star$, see text for details.}
\end{tablenotes}
\end{threeparttable}
\end{table*}

It is valuable to consider how strong the signal from our targets could be, given EM observations. If we assume that all rotational energy losses, as observed in the frequency derivative $\dot{f}_\star$, are converted into gravitational radiation, the indirect spin-down limit on the maximum strain, $h_{0,\,\textrm{sd}}$, is \cite{Riles2013a}
\begin{align}
h_{0,\,\textrm{sd}} = &~4.0\times 10^{-28} \left( \frac{8\,{\rm kpc}}{D}\right) \nonumber \\
					&\times \left(\frac{600\,{\rm Hz}}{f_{\rm GW}} \right)^{1/2} \left(\frac{-\dot{f}_{\rm GW}}{10^{-14}\,{\rm Hz\,s}^{-1}} \right)^{1/2}\ , \label{eq:h0sd}
\end{align}
where $D$ is the distance to the target, $f_{\rm GW}$ is the gravitational wave frequency, and $\dot{f}_{\rm GW}$ is its derivative. In Eq.~\eqref{eq:h0sd} we assume $I_{zz} / I_0 \approx 1$, i.e.~the $zz$ component of the moment-of-inertia tensor ($I_{zz}$) is very close to the moment-of-inertia of an undeformed star ($I_0$). We assume $f_{\rm GW} \approx 2 f_\star$ when computing Eq.~\eqref{eq:h0sd} for each of our targets. We list the best estimates for the distance to each target in the second column of Table \ref{tab:exp_ul}. These estimates are typically poorly known, especially if there is no known counterpart observed in wavelengths other than X-ray for the target. We use the central estimate of the distance in Eq.~\eqref{eq:h0sd}.

For AMXPs, $\dot{f}_\star$ is estimated by constructing a phase-connected timing solution when the target is in outburst, but estimates for $\dot{f}_\star$ in quiescence are also possible for targets that have gone into outburst multiple times. The $\dot{f}_\star$ observed during outburst can be either positive (corresponding to spin-up) or negative (corresponding to spin-down), while in quiescence $\dot{f}_\star$ is typically (but not always) negative \cite{Ghosh1977, Melatos2016}. The third column of Table \ref{tab:exp_ul} records $\dot{f}_\star$ for each of our targets. When $\dot{f}_\star$ has been measured in multiple outburst events, only the $\dot{f}_\star$ from the most recent outburst is listed. For $\dot{f}_\star < 0$ we assume $\dot{f}_\textrm{GW} \approx 2 \dot{f}_\star$ in Eq.~\eqref{eq:h0sd}. For targets with $\dot{f}_\star < 0$ (in either quiescent or active phases) we find $10^{-28} \lesssim h_{0,\,\textrm{sd}} \lesssim 10^{-27}$ (fourth column of Table \ref{tab:exp_ul}), an order of magnitude lower than the estimated value of $\hul$. 

As argued in Ref.~\cite{Middleton2020}, for $\dot{f}_\star > 0$ the torque due to gravitational radiation reaction may be masked by the accretion torque, allowing larger values of $\dot{f}_\textrm{GW}$, as long as one has $\dot{f}_\star = \dot{f}_\textrm{acc} + \dot{f}_\textrm{GW}$, where $\dot{f}_\textrm{acc}$ is the spin-up rate due to accretion. A reasonable choice, without excessive fine-tuning, is to set $\dot{f}_\textrm{GW} \approx - \dot{f}_\star$, for an order-of-magnitude estimate in Eq.~\eqref{eq:h0sd}, i.e.~assuming $|\dot{f}_\textrm{acc}| \approx 2 |\dot{f}_\textrm{GW}|$. The resultant values for $h_{0,\,\textrm{sd}}$ for targets with $\dot{f}_\star > 0$ are all well below the estimates of $\hul$ set in Sec.~\ref{sec:ul_results}, and fall in the range $10^{-28} \lesssim h_{0,\,\textrm{sd}} \lesssim 10^{-27}$. 

Another avenue through which EM observations can constrain $h_0$ is by assuming that the X-ray flux is proportional to the mass accretion rate, and that the torque due to accretion balances the gravitational radiation reaction. The torque-balance limit is \cite{Riles2013a, Zhang2021}
\begin{align}
h_{0,\,\textrm{torque}} = &~5\times 10 ^{-27}\left(\frac{600\,\textrm{Hz}}{f_\textrm{GW}}\right)^{1/2} \nonumber \\
						&\times \left(\frac{F_X}{10^{-8}\,\textrm{erg\,s$^{-1}\,$cm$^{-2}$}}\right)^{1/2}\ ,  \label{eq:h0t}
\end{align}
where $F_X$ is the observed bolometric X-ray flux. Eq.~\eqref{eq:h0t} has a few hidden assumptions, namely: \begin{enumerate*}[label=\roman*)]
\item that the mass of the neutron star is $1.4M_\odot$,
\item that all of the accretion luminosity is radiated as an X-ray flux, and
\item that the accretion torque is applied at the radius of the neutron star, which is set to 10\,km.
\end{enumerate*}
The exact dependence of the torque-balance limit on these assumptions is discussed in Ref.~\cite{Zhang2021}. We take $f_{\rm GW} \approx 2 f_\star$ for each of our targets, as for Eq.~\eqref{eq:h0sd}. We take $F_X = F_{X,\,\textrm{max}}$, the maximum recorded X-ray flux from each target when it was in outburst (fifth column of Table \ref{tab:exp_ul}), providing an upper limit on $h_{0,\,\textrm{torque}}$ (sixth column of Table \ref{tab:exp_ul}). We find $5\times 10^{-28} \lesssim h_{0,\,\textrm{torque}} \lesssim 1\times 10^{-27}$ across all targets. 

\subsection{Astrophysical implications \label{sec:ul_impl}}

\begin{figure*}
	\centering
	\includegraphics[width=0.8\linewidth]{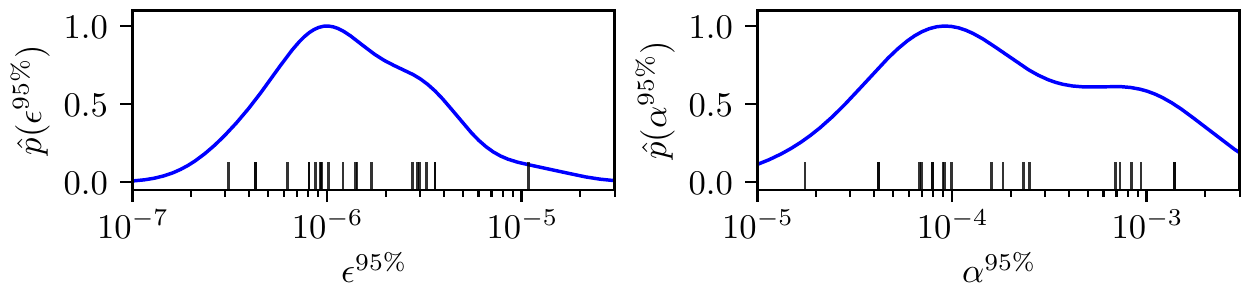}
	\caption{Kernel density estimate of the PDF of the constraints on ellipticity $\epsilon^{95\%}$ (left panel) and dimensionless $r$-mode amplitude $\alpha^{95\%}$ (right panel) via Eqs.~\eqref{eq:ellip} and \eqref{eq:alpha} respectively. Both PDFs are normalized to a height of one. The black dashes in both panels correspond to the individual estimates of $\epsilon^{95\%}$ or $\alpha^{95\%}$ from each target.}
	\label{fig:kde}
\end{figure*}

The estimates of $\hul$ given in Sec.~\ref{sec:ul_results} can be converted into constraints on the physical parameters that govern the mechanism putatively generating continuous gravitational waves in each sub-band.

In the $2f_\star$ sub-band the simplest emission mechanism is that of a perpendicular biaxial rotator (using the language from Ref.~\cite{Sun2019}), for which we calculate the upper limit of the ellipticity of the neutron star as \cite{Jaranowski1998}
\begin{equation}
\epsilon^{95\%} = 2.1 \times 10^{-6} \left(\frac{\hul}{10^{-25}} \right) \left(\frac{D}{8\,\textrm{kpc}} \right) \left(\frac{600\,\textrm{Hz}}{f_{\textrm{GW}}} \right)^2\ , \label{eq:ellip}
\end{equation}
assuming $I_{zz} = 10^{38}\,$kg\,m$^2$. Using the central estimate for $D$ (second column of Table \ref{tab:exp_ul}), we find the strictest constraint, from all of our targets, $\epsilon^{95\%} = 3.1\times 10^{-7}$ for IGR J00291$+$5934. A kernel density estimate of the probability density function (PDF) of the constraints $\epsilon^{95\%}$, $\hat{p}(\epsilon^{95\%})$, for all our targets, is shown in the left panel of Fig.~\ref{fig:kde}. It is peaked around $\epsilon^{95\%} \sim 10^{-6}$. 

In the $4f_\star/3$ sub-band the emission mechanism is via $r$-modes, the strength of which is parametrized as \cite{Owen2010}
\begin{equation}
\alpha^{95\%} = 1.0\times10^{-4} \left(\frac{\hul}{10^{-25}} \right) \left(\frac{D}{8\,\textrm{kpc}} \right) \left(\frac{600\,\textrm{Hz}}{f_{\textrm{GW}}} \right)^3\ . \label{eq:alpha}
\end{equation}
Eq.~\eqref{eq:alpha} assumes $f_{\textrm{GW}} \approx 4f_\star/3$, which may not be true, as discussed in Sec.~\ref{sec:params} \cite{Idrisy2015, Caride2019}. The strictest constraint, from all of our targets, is $\alpha^{95\%} = 1.8\times 10^{-5}$, again for IGR J00291$+$5934. A kernel density estimate of the PDF of the constraints $\alpha^{95\%}$, $\hat{p}(\alpha^{95\%})$, for all our targets, is shown in the right panel of Fig.~\ref{fig:kde}. It is peaked around $\alpha^{95\%} \sim 10^{-4}$.

The kernel density estimates of the PDFs $\hat{p}(\epsilon^{95\%})$ and $\hat{p}(\alpha^{95\%})$ in Fig.~\ref{fig:kde} are not constraints on $\epsilon$ and $\alpha$ respectively, nor are they expressing the uncertainty in each individual estimate of $\epsilon^{95\%}$ or $\alpha^{95\%}$ (which are dominated by the uncertainty in $\hul$, and the distance, see column two of Table~\ref{tab:exp_ul}). They are instead presented to indicate where the constraints on $\epsilon^{95\%}$ and $\alpha^{95\%}$ lie, given the strain upper limits calculated for the targets in this search. That is, they are estimates of the true probability distribution of the constraints one would obtain for $\epsilon$ and $\alpha$, given a large population of AMXPs (assuming the targets studied here are representative of this larger population). The kernel density estimates are calculated by summing Gaussian kernels centered on each data point, with bandwidth chosen to minimize the asymptotic mean integrated square error \cite{Wand1995}.

The physical mechanism for emission in the $f_\star$ sub-band is less well-defined. A biaxial non-perpendicular rotator emits gravitational radiation at both $f_\star$ and $2f_\star$ \cite{Jaranowski1998, Jones2010, Jones2015}. The emission at $f_\star$ dominates the $2f_\star$ emission for both $\theta \lesssim 20\degr$ and $|\cos\iota| \lesssim 0.8$, where $\theta$ is the wobble angle (see figure 5 of Ref.~\cite{Sun2019} for details). The value of $\theta$ is low for certain models involving pinned superfluid interiors \cite{Jones2010, Melatos2015}. Other possibilities exist, including a triaxial rotator \cite{Zimmermann1980, Broeck2005, Lasky2013}. We recommend future searches to also consider searching the $f_\star$ sub-band, due to the wealth of information that a continuous gravitational wave detection at this frequency would provide regarding neutron star structure.

\section{Conclusions} \label{sec:concl}
We present the results of a search for continuous gravitational waves from 20 accreting low-mass X-ray binaries in the Advanced LIGO O3 dataset. Five of these targets were searched before in O2 \cite{Middleton2020}, and one was searched in S6 \cite{s6twoSpectScoXTE}. The search pipeline we use allows for spin-wandering and tracks the orbital phase of the binary via a hidden Markov model and the $\mj$-statistic respectively. The targets have well-constrained rotational frequencies, $f_\star$, and orbital elements from electromagnetic observations of outburst events, restricting the parameter space. For each target we search three $\sim0.61\,$Hz-wide sub-bands centered on $\{1, 4/3, 2\}f_\star$. We also perform a target-of-opportunity search for emission from \sax, which went into outburst during O3a.

We find no candidates that survive our veto procedure and are above a threshold corresponding to a 1\% false alarm probability per sub-band. We find 16 candidates that survive our astrophysical vetoes when we set the threshold to 30\% false alarm probability per sub-band. As we search a total of 60 sub-bands, this number of surviving candidates is consistent with the expected number of false alarms. These candidates are systematically investigated with further follow-up. In all cases, the follow-up does not provide convincing evidence that any are real astrophysical signals. However, they could not be convincingly ruled out, which is not surprising given their borderline significance. We record the orbital template and frequencies recovered for these candidates, and recommend that they are followed up in future gravitational wave data sets, and with different pipelines. 

The target-of-opportunity search returns one candidate above threshold that survives our veto procedure. Additional, detailed follow-up of this candidate does not produce convincing evidence that it is a true astrophysical signal rather than a noise fluctuation.

Assuming all of the candidates are not astrophysical, we set upper limits on the strain at 95\% confidence in each sub-band. Using these estimates, the strictest constraint on neutron star ellipticity is $\epsilon^{95\%} = 3.1\times 10^{-7}$. The strictest constraint we place on the $r$-mode amplitude is $\alpha^{95\%} = 1.8\times 10^{-5}$. Both of these constraints come from IGR J00291$+$5934.

\section*{Acknowledgments}
This material is based upon work supported by NSF’s LIGO Laboratory which is a major facility
fully funded by the National Science Foundation.
The authors also gratefully acknowledge the support of
the Science and Technology Facilities Council (STFC) of the
United Kingdom, the Max-Planck-Society (MPS), and the State of
Niedersachsen/Germany for support of the construction of Advanced LIGO 
and construction and operation of the GEO600 detector. 
Additional support for Advanced LIGO was provided by the Australian Research Council.
The authors gratefully acknowledge the Italian Istituto Nazionale di Fisica Nucleare (INFN),  
the French Centre National de la Recherche Scientifique (CNRS) and
the Netherlands Organization for Scientific Research, 
for the construction and operation of the Virgo detector
and the creation and support  of the EGO consortium. 
The authors also gratefully acknowledge research support from these agencies as well as by 
the Council of Scientific and Industrial Research of India, 
the Department of Science and Technology, India,
the Science \& Engineering Research Board (SERB), India,
the Ministry of Human Resource Development, India,
the Spanish Agencia Estatal de Investigaci\'on,
the Vicepresid\`encia i Conselleria d'Innovaci\'o, Recerca i Turisme and the Conselleria d'Educaci\'o i Universitat del Govern de les Illes Balears,
the Conselleria d'Innovaci\'o, Universitats, Ci\`encia i Societat Digital de la Generalitat Valenciana and
the CERCA Programme Generalitat de Catalunya, Spain,
the National Science Centre of Poland and the Foundation for Polish Science (FNP),
the Swiss National Science Foundation (SNSF),
the Russian Foundation for Basic Research, 
the Russian Science Foundation,
the European Commission,
the European Regional Development Funds (ERDF),
the Royal Society, 
the Scottish Funding Council, 
the Scottish Universities Physics Alliance, 
the Hungarian Scientific Research Fund (OTKA),
the French Lyon Institute of Origins (LIO),
the Belgian Fonds de la Recherche Scientifique (FRS-FNRS), 
Actions de Recherche Concertées (ARC) and
Fonds Wetenschappelijk Onderzoek – Vlaanderen (FWO), Belgium,
the Paris \^{I}le-de-France Region, 
the National Research, Development and Innovation Office Hungary (NKFIH), 
the National Research Foundation of Korea,
the Natural Science and Engineering Research Council Canada,
Canadian Foundation for Innovation (CFI),
the Brazilian Ministry of Science, Technology, and Innovations,
the International Center for Theoretical Physics South American Institute for Fundamental Research (ICTP-SAIFR), 
the Research Grants Council of Hong Kong,
the National Natural Science Foundation of China (NSFC),
the Leverhulme Trust, 
the Research Corporation, 
the Ministry of Science and Technology (MOST), Taiwan,
the United States Department of Energy,
and
the Kavli Foundation.
The authors gratefully acknowledge the support of the NSF, STFC, INFN and CNRS for provision of computational resources.
This work was supported by MEXT, JSPS Leading-edge Research Infrastructure Program, JSPS Grant-in-Aid for Specially Promoted Research 26000005, JSPS Grant-in-Aid for Scientific Research on Innovative Areas 2905: JP17H06358, JP17H06361 and JP17H06364, JSPS Core-to-Core Program A. Advanced Research Networks, JSPS Grant-in-Aid for Scientific Research (S) 17H06133, the joint research program of the Institute for Cosmic Ray Research, University of Tokyo, National Research Foundation (NRF) and Computing Infrastructure Project of KISTI-GSDC in Korea, Academia Sinica (AS), AS Grid Center (ASGC) and the Ministry of Science and Technology (MoST) in Taiwan under grants including AS-CDA-105-M06, Advanced Technology Center (ATC) of NAOJ, and Mechanical Engineering Center of KEK. 

This work is supported by NASA through the NICER mission and the Astrophysics Explorers Program and uses data and software provided by the High Energy Astrophysics Science Archive Research Center (HEASARC), which is a service of the Astrophysics Science Division at NASA/GSFC and High Energy Astrophysics Division of the Smithsonian Astrophysical Observatory.

\appendix
\section{Threshold setting \label{app:thresh}}
In this Appendix we outline two alternative methods to set thresholds for the search. In Appendix \ref{app:exp} we detail the method in Ref.~\cite{o2vitsco} to set thresholds by modeling the tail of the log-likelihood distribution in noise as an exponential. In Appendix \ref{app:perc} we review the non-parametric method in Refs.~\cite{o1vitsco, Middleton2020, Millhouse2020, Jones2021, Beniwal2021}, which takes a certain percentile detection statistic from noise-only realizations as the threshold. We compare the methods in Appendix \ref{app:whyth}. In Appendix \ref{app:offt_th} we discuss generating noise realizations using off-target searches, and justify the approach taken in this paper. In Appendix \ref{app:pnoise} we specify how to calculate $\pn$, the probability that we see a value of $\ml$ at least as high as a certain candidate in a given sub-band.

Whatever the method, the threshold depends on both the target's projected semi-major axis, $a_0$, and the sub-band frequency, $f$, as log-likelihoods depend non-linearly on $a_0\, f$ as an increased number orbital sidebands are included in the $\mj$-statistic at higher $a_0\, f$ [see equation (6) in Ref.~\cite{o2vitsco} and Ref.~\cite{Suvorova2017} for details]. For this reason we set thresholds independently for each target and sub-band. 

\subsection{Exponential tail method  \label{app:exp}}
The PDF of the log-likelihood, $p(\ml)$, for the most likely path for a given template is observed to have an exponentially distributed tail in noise,
\begin{equation}
p(\ml) = A \lambda \exp \left[-\lambda \left( \ml - \ml_{\rm tail}\right) \right] \quad {\rm for} \quad \ml > \ml_{\rm tail}\ , \label{eq:pl}
\end{equation}
where $A$ is a normalization constant, $\lambda$ is a parameter to be found empirically, and $\ml_{\rm tail}$ is a cut-off that must also be determined empirically. 

For each target and sub-band we estimate $\lambda$ and $\ml_{\rm tail}$ using a set of $M$ sample log-likelihoods, a subset of which have $\ml > \ml_{\rm tail}$. This subset is denoted $S_{N_{\rm tail}} \equiv \{\ml_i \}$, $i \in \{1,...,N_{\rm tail} \}$. The entire set of $M$ samples is generated by running the search on $N_{\rm G} = 100$ realizations of Gaussian noise. To keep $N_{\rm G}$ small enough to be computationally feasible we include log-likelihoods from all possible Viterbi paths through the sub-band for each template, instead of just the log-likelihood from the most likely path. Thus, we have $M = N_{\rm G} N_f N_B$, where $N_f = 2^{20}$ is the number of frequency bins in each sub-band, and $N_B$ is the number of binary orbital templates needed for each individual sub-band, as listed in Table \ref{tab:ntemps}. Separate tests, not shown here, indicate that including the log-likelihoods from non-maximal paths does not change the shape of $p(\ml)$, and therefore does not change the thresholds $\lth$, if the appropriate trials factor is taken into account.

Assuming each $\ml_i$ is independent, the maximum likelihood estimator, $\hat{\lambda}$, for $\lambda$ is
\begin{equation}
\hat{\lambda} = \frac{N_{\rm tail}}{\sum_{i=1}^{N_{\rm tail}} \left(\ml_i - \ml_{\rm tail} \right)}\ .
\end{equation}
The normalization $A = N_{\rm tail} / M$ is fixed via the fraction of total samples used to construct $p(\ml)$. The cut-off $\ml_{\rm tail}$ is estimated in each sub-band as the smallest value $\ml^*$ where a histogram of the samples $\ml_i > \ml^*$ has approximately constant slope when viewed on log-linear axes. Each $\ml_i$ is independent for the long coherence times ($\tdr=10\,$d) used in this search, as $N_T \ll N_f$ implies most optimal paths through the sub-band are not correlated. 

The probability, $\alpha$, that $\ml$ is above some threshold $\lth > \ml_{\rm tail}$ if no signal is present (i.e.~in pure noise) is
\begin{equation}
\int_{\lth}^{\infty} {\rm d}\ml\, p(\ml) = \alpha\ . \label{eq:palpha}
\end{equation}
Combining Eqs.~\eqref{eq:alphan}, \eqref{eq:pl}, and \eqref{eq:palpha} we solve for $\lth$ in a given sub-band, viz.~
\begin{equation}
\lth = -\frac{1}{\hat{\lambda}} \log \left(\frac{N_{\rm G} \alpha_{N_{\rm tot}}}{N_{\rm tail}}\right) + \ml_{\rm tail}\ , \label{eq:lth}
\end{equation}
where Eq.~\eqref{eq:alphan} is simplified via the binomial approximation ($N_{\rm tot} = N_f N_B \gg 1$), and $\alpha_{N_{\rm tot}}$ is the desired false alarm probability for the search over the sub-band. Note Eq.~\eqref{eq:lth} depends implicitly on the sub-band frequency and $N_B$, through $\hat{\lambda}$ and $N_{\rm tail}$. 

Across all targets and sub-bands we find $0.195 < \hat{\lambda} < 0.248$, with larger values corresponding to higher frequency sub-bands, and those with larger $N_B$. A simple rule-of-thumb is that, for a median value of $\hat{\lambda}=0.218$, an increment of $\approx 3$ in $\ml$ is $\approx 50\%$ less likely to occur in pure noise.

\subsection{Percentile method \label{app:perc}}
Given a sorted set of most likely log-likelihoods $\{\ml_i\}$, $i \in \{1, ..., M\}$ with $M = N_{\rm G} N_B$, generated via running the search algorithm over $N_{\rm G}$ realizations of noise for a single target and sub-band, one can pick as the threshold the $\ml_i$ corresponding to the percentile equal to the desired false alarm probability, i.e.
\begin{equation}
\lth = \ml_{j}\ , \label{eq:lth_perc}
\end{equation}
with $j = \lfloor \alpha_{N_{\rm tot}} M \rfloor$. As with the method described in Appendix \ref{app:exp} we may opt to use the log-likelihoods from all possible Viterbi paths through the sub-band for a given orbital template, to reduce the number of realizations of noise we need to generate. With this set of log-likelihoods, we have $M=N_{\rm G}  N_f  N_B$. 

\subsection{Comparison of methods \label{app:whyth}}
The two methods described in Appendices \ref{app:exp} and \ref{app:perc} give broadly similar results for $\lth$ for a given probability of false alarm. Ref.~\cite{o2vitsco} opts for the method in Appendix \ref{app:exp}. When Viterbi scores are used as the detection statistic, as in Ref.~\cite{o2vitsco}, the PDF of the score in noise does not vary with frequency, and thus the thresholds in each sub-band can be extrapolated from a small set of Gaussian noise realizations. If the PDF of the detection statistic varies with target search parameters, then the method in Appendix \ref{app:perc} is used, as in Refs.~\cite{Middleton2020, Millhouse2020, Jones2021, Beniwal2021}. The percentile method has inherently fewer assumptions, as it does not fit a parametric model to $p(\ml)$. However it is not possible to extrapolate thresholds calculated in one sub-band to other sub-bands. 

For our targets and sub-bands, we find $\mathcal{L}_{\rm th}^e - \mathcal{L}_{\rm th}^p \approx 2$, where the superscripts $e$ and $p$ correspond to the exponential tail and percentile methods respectively. However the exact difference depends on the realizations of Gaussian noise; Monte Carlo simulations indicate that with $N_{\rm G} = 100$ the calculated threshold is usually within 2\% of the true value, so thresholds should only be considered precise to 2\%. 

\subsection{Off-target thresholds \label{app:offt_th}}
Both methods derive $\lth$ based on realizations of Gaussian noise. However, the noise in real detector data is non-Gaussian in general \cite{o23DetChar}. To account for this we search O3 data at $N_{\rm OT}$ randomly chosen, but well-separated, off-target positions, to generate $N_{\rm OT}$ realizations of real detector noise, as originally done in Ref.~\cite{Middleton2020}. We set $N_{\rm OT}$ such that $N_B N_{\rm OT} > 500$, with a minimum value of $N_{\rm OT}=100$, to ensure enough samples are generated.

If there are no known noise lines in the sub-band, we find $4 < \lthg^e - \lthot^p < 12$, where the subscripts G and OT correspond to thresholds calculated using Gaussian and off-target noise realizations respectively. That is, the thresholds calculated from Gaussian noise, using the exponential tail method are considerably more conservative than those calculated from off-target noise and the percentile method. If there are loud noise lines in the sub-band, $\lthot$ is often much higher, as these lines appear in the off-target noise realizations. Because off-target noise realizations are impacted by noise lines, $p(\ml)$ is not necessarily exponential in its tail. We thus opt to use the percentile method when calculating thresholds with off-target noise realizations. Table \ref{tab:threshes} contains the calculated $\lthg^e$ and $\lthot^p$ for each target and sub-band.

As in Ref.~\cite{Middleton2020} we consider $\lth$ for each sub-band to be the minimum of $\lthg^e$ and $\lthot^p$, with $\alpha_{N_{\rm tot}}=0.3$. This choice minimizes the probability that we will miss a potential candidate due to inadvertently setting our threshold too high. 

\subsection{Probability that a candidate arises due to noise \label{app:pnoise}}
As discussed in Sec.~\ref{sec:thresh}, when we set $\alpha_{N_{\rm tot}} = 0.3$ we expect $\sim18$ candidates above $\lth$, across all targets and sub-bands. Let us quantify empirically the probability, $\pn$, that, if the data in a given sub-band are pure noise, we see at least one template with log-likelihood higher than that of the candidate, $\ml_{\rm cand}$. We have
\begin{equation}
\pn = \frac{\sum_{i=1}^M \mathbbm{1}\left(\ml_i > \ml_{\rm cand} \right)}{M}\ , \label{eq:pn}
\end{equation}
where $\mathbbm{1}(...)$ is the indicator function which returns 1 when the argument is true, otherwise 0. In this paper we calculate Eq.~\eqref{eq:pn} for each candidate with $\ml > \lth$ using the set of log-likelihoods, $\{\ml_i\}$, generated via off-target realizations as discussed in Appendix \ref{app:offt_th}. As in Appendix \ref{app:perc}, we set $M = N_{\rm G} N_B$ to account for the extra ``trials factor'' needed for sub-bands with multiple templates.

\begin{table*}
\caption{Target, starting frequency, $f_{\rm s}$, for each $\sim0.61\,$Hz-wide sub-band, threshold calculated using Gaussian noise realizations and the exponential tail method, $\lthg^{e}$, and threshold calculated using off-target noise realizations and the percentile method, $\lthot^{p}$. All thresholds are calculated with $\alpha_{N_{\rm tot}}=0.3$. \label{tab:threshes}}
\begin{ruledtabular}
\begin{tabular}{l d{4} d{2} d{2} l d{4} d{2}  d{2}}
\textrm{Target} & \textrm{$f_{\rm s}$ (Hz)} & \textrm{$\lthg^{e}$} & \textrm{$\lthot^{p}$} & \textrm{Target} & \textrm{$f_{\rm s}$ (Hz)} & \textrm{$\lthg^{e}$} & \textrm{$\lthot^{p}$} \\
\midrule
IGR J00291$+$5934	& 598.6		& 291.9 & 1136.7	 &  IGR J17498$-$2921	& 400.7		& 304.6 & 298.4 	\\ 
 					& 798.2		& 294.9 & 288.4 	 &   					& 534.4		& 304.7 & 297.9 	\\ 
 					& 1197.5	& 295.2 & 287.6 	 &   					& 801.7		& 311.2 & 304.5 	\\ 
MAXI J0911$-$655	& 339.7		& 297.0 & 290.0      &  IGR J17511$-$3057	& 244.5		& 302.0 & 293.7 	\\ 
	 				& 453.0		& 305.4 & 298.2 	 &   					& 326.1		& 303.4 & 295.8 	\\ 
 	   				& 679.6		& 305.5 & 300.4 	 &   					& 489.4		& 305.1 & 297.9 	\\ 
XTE J0929$-$314		& 184.8		& 311.9 & 304.9 	 &  XTE J1751$-$305		& 435.0		& 312.8 & 316.0 	\\ 
					& 246.5		& 307.4 & 301.2 	 &  					& 580.1		& 312.9 & 306.6 	\\ 
					& 369.9		& 310.4 & 304.5 	 &  					& 870.3		& 319.9 & 315.7 	\\ 
IGR J16597$-$3704	& 104.9		& 321.6 & 316.4 	 &  Swift J1756.9$-$2508& 181.8		& 308.7 & 302.8     \\ 
 					& 139.9		& 322.9 & 625.5 	 &  					& 242.5		& 317.4 & 312.8 	\\ 
 					& 210.0		& 323.7 & 318.5 	 &  					& 363.8		& 315.3 & 309.0 	\\ 
IGR J17062$-$6143	& 163.4		& 292.1 & 285.1 	 &  IGR J17591$-$2342	& 527.1		& 299.3 & 289.4 	\\ 
 					& 217.9		& 289.1 & 281.8 	 &   					& 702.9		& 302.0 & 295.7 	\\ 
 					& 327.0		& 293.3 & 283.3 	 &   					& 1054.5	& 304.7 & 298.2 	\\ 
IGR J17379$-$3747	& 467.8		& 307.4 & 298.9 	 &  SAX J1808.4$-$3658	& 400.7		& 303.8 & 294.8 	\\ 
 					& 623.8		& 307.4 & 299.9 	 &  					& 534.3		& 305.0 & 296.1 	\\ 
 					& 935.9		& 311.1 & 305.7 	 &  					& 801.6		& 309.3 & 301.4 	\\ 
SAX J1748.9$-$2021	& 442.1		& 308.2 & 300.3 	 &  XTE J1807$-$294		& 190.3		& 295.5 & 287.0 	\\ 
					& 589.5		& 310.1 & 301.6 	 &  					& 253.9		& 296.8 & 289.0 	\\ 
					& 884.4		& 311.5 & 304.9 	 &  					& 380.9		& 299.7 & 292.0 	\\ 
NGC 6440 X$-$2		& 205.6		& 292.8 & 281.3      &  XTE J1814$-$338		& 314.1		& 301.8 & 293.3 	\\ 
					& 274.2		& 298.3 & 288.6 	 &  					& 418.8		& 302.3 & 294.3 	\\ 
					& 411.5		& 295.9 & 287.2 	 &  					& 628.4		& 305.8 & 298.4 	\\ 
IGR J17494$-$3030	& 375.7		& 315.3 & 309.3 	 &  IGR J18245$-$2452	& 254.0		& 311.2 & 305.4 	\\ 
 					& 501.1		& 317.5 & 13763.8 	 &   					& 338.8		& 312.5 & 305.9 	\\ 
 					& 751.8		& 322.2 & 316.5 	 &   					& 508.4		& 317.3 & 13569.7 	\\ 
Swift J1749.4$-$2807& 517.6		& 316.4 & 308.8      &  HETE J1900.1$-$2455	& 377.0		& 299.6 & 288.1 	\\ 
					& 690.3		& 318.4 & 311.9 	 &   					& 502.8		& 303.1 & 8459.4 	\\ 
					& 1035.5	& 321.0 & 316.3 	 &   					& 754.3		& 303.8 & 294.7 	\\                                                                                           
\end{tabular}
\end{ruledtabular}
\end{table*}

\section{ \label{app:fullresults} Full search results and survivor follow-up}
This Appendix collates the full search results for reference and reproducibility for all targets in Figs.~\ref{fig:igra}--\ref{fig:hete} (except for \igrh\ which is shown in Fig.~\ref{fig:igrh}). Each of Figs.~\ref{fig:igra}--\ref{fig:hete} is laid out identically to Fig.~\ref{fig:igrh}. 

\begin{table*}
\caption{Orbital template, $(P,\ a_0,\ \tasc)$, terminating frequency bin, $f(N_T)$, log-likelihood, $\ml$, and the probability that a search of the candidate's sub-band in pure noise would return a candidate just as loud, $\pn$, for the 16 candidates with $\ml > \lth$ that cannot be eliminated by any of the vetoes detailed in Sec.~\ref{sec:vetoes}. \label{tab:outliers}}
\begin{ruledtabular}
\begin{tabular}{l c r r r r r r}
\textrm{Target} & \textrm{Candidate} & \textrm{$P$ (s)} & \textrm{$a_0$ (lt-s)} & \textrm{$\tasc$ (GPS time)} & \textrm{$f(N_T)$ (Hz)} & \textrm{$\ml$} & \textrm{$\pn$} \\
\midrule
MAXI J0911$-$655    & 1 &  2659.933		& 0.0176 	 & 1238165869.0437 & 453.309532 & 299.2 & 0.26   	\\
IGR J16597$-$3704	& 1 &  2758.61		& 0.0048	 & 1238163275.6122 & 105.002195 & 316.5 & 0.30		\\
					& 2 &  2757.90		& 0.0048	 & 1238163010.7583 & 210.359055 & 323.5 & 0.09		\\
IGR J17062$-$6143   & 1 &  2278.2112	& 0.0040	 & 1238165942.2745 & 163.531805 & 286.4 & 0.24 		\\
	    			& 2 &  2278.2112	& 0.0040	 & 1238165942.2745 & 218.452091 & 283.9 & 0.19 		\\
	    			& 3 &  2278.2112	& 0.0040	 & 1238165942.2745 & 327.058287 & 290.0 & 0.05 		\\
IGR J17379$-$3747   & 1 &  6765.84		& 0.0770	 & 1238162768.3832 & 623.819568 & 303.9 & 0.08    	\\
SAX J1748.9$-$2021  & 1 &  31555.29		& 0.3876	 & 1238151700.2214 & 590.048237 & 304.9 & 0.12    	\\
	    			& 2 &  31555.30		& 0.3876	 & 1238151760.9764 & 590.040010 & 302.3 & 0.27    	\\
	    			& 3 &  31555.31		& 0.3876	 & 1238151710.6406 & 884.592276 & 305.6 & 0.22    	\\
IGR J17498$-$2921   & 1 &  13835.619	& 0.36517	 & 1238164013.8774 & 801.703605 & 305.8 & 0.22 		\\
XTE J1807$-$294   	& 1 &  2404.416		& 0.00483	 & 1238165585.2721 & 381.000852 & 296.7 & 0.10   	\\
SAX J1808.4$-$3658  & 1 &  7249.15		& 0.0628	 & 1238161168.0040 & 534.633578 & 298.2 & 0.16     	\\
	    			& 2 &  7249.16		& 0.0628	 & 1238161183.0831 & 534.407934 & 296.2 & 0.30     	\\
XTE J1814$-$338   	& 1 &  15388.723	& 0.3906	 & 1238151585.3941 & 314.564137 & 297.7 & 0.08   	\\
HETE J1900.1$-$2455 & 1 &  4995.26		& 0.0184	 & 1238161529.0866 & 754.378543 & 295.8 & 0.25   	\\
\end{tabular}
\end{ruledtabular}
\end{table*}

The orbital parameters ($P$, $a_0$, and $\tasc$), terminating frequency bin [$f(N_T)$], log-likelihood ($\ml$), and $\pn$, the probability that a search of that candidate's sub-band in pure noise would return at least one candidate at least as loud as the one seen are shown in Table \ref{tab:outliers}, for each of the candidates that survive all vetoes and have $\ml > \lth$. 

\subsection{Additional follow-up for survivors \label{app:followup}}
The full frequency paths, $f(t) - f(N_T)$, for all candidates with $\pn \leq 0.1$ are shown in the top panels of Figs.~\ref{fig:igrb_f}--\ref{fig:xted_f}. The bottom panels of Figs.~\ref{fig:igrb_f}--\ref{fig:xted_f} display the cumulative log-likelihood along the frequency path relative to the average sum log-likelihood needed to reach $\lth$, namely $C\ml \equiv \sum_{i=0}^{i=t}\big[\ml(i) - \lth / N_T\big]$, where $\sum_{i=0}^{i=t}\ml(i)$ is ${\rm ln\,} P(Q^*|O)$ from Eq.~\eqref{eq:probq_o} truncated after the $t$-th segment. Over-plotted (blue dashed line) is the average cumulative log-likelihood needed at each data segment in order to reach $\lth$. This diagnostic indicates whether a handful of segments dominate in making the candidate's frequency path the optimal one for that template. If the candidate is a true signal, we would expect the signal strength to be approximately constant, and thus the cumulative log-likelihood should grow linearly as more data are considered. However, Monte Carlo tests with injections show that the cumulative log-likelihood only becomes linear for $\ml \gtrsim \lth+200$. This is not the case for any of the 16 survivor candidates, and thus their cumulative log-likelihood cannot help us distinguish whether they are truly astrophysical signals.

The sky resolution of the algorithm described in Sec.~\ref{sec:alg} is roughly 2\, arcmin in RA and Dec., for an injection with $\lth \lesssim \ml \lesssim \lth + 50$. The point-spread-function of an injection is an ellipse, which has a varying orientation and eccentricity dependent on the sky position. For each of our candidates we calculate $\ml$ at 440 regularly spaced sky positions in a 100\,arcmin$^2$ grid around the target's true location, using the template recovered from the search and listed in Table \ref{tab:outliers}. For almost all survivor candidates, the distribution of $\ml$ values in the patch of sky around the candidate does not match the elliptical point-spread-function we see in injections for their respective sky locations. The sole exception is Candidate 2 from IGR J16597$-$3704. Figure \ref{fig:skymap} shows $\ml$ at 3721 regularly spaced sky positions in a 100\,arcmin$^2$ grid around the target's true location, again using the template as listed in Table \ref{tab:outliers}. The roughly elliptical shape is consistent with the point-spread-function of injections at this sky location. However, the region of sky with $\ml \gtrsim \lth$ is centered $\sim 1\,$arcmin lower in Dec. than the true declination of the source, which is known to a precision of 0.01\,arcmin \cite{Tetarenko2018}. 

One final follow-up we perform for these candidates is to calculate $\ml$ in a small, densely sampled patch of the $\{P,\ \tasc\}$ parameter space around each candidate's template. Moderately loud injections ($\ml \gtrsim \lth + 100$) are seen to ``spread out'' in the $\{P,\ \tasc\}$ plane, and are detectable with $\ml > \lth$ even when searching a template that has a slightly incorrect value of $P$ and $\tasc$. However, none of our candidates are this loud, so this diagnostic does not help us distinguish whether they are truly astrophysical signals or merely noise fluctuations.

We do not use any data from LIGO's Observing Runs 1 or 2 (O1 and O2 respectively) to aid in following up these candidates, as the detector is considerably more sensitive in O3. The duration of O3 was also longer than the durations of O1 and O2. If a candidate is only marginally above threshold in O3 data, it may be hidden in the noise in O1 and O2 data, so including data from those observing runs is not likely to increase the candidate's signal-to-noise ratio. 

\begin{figure*}
	\centering
	\begin{subfigure}[t]{0.8\textwidth}
		\includegraphics[width=\linewidth]{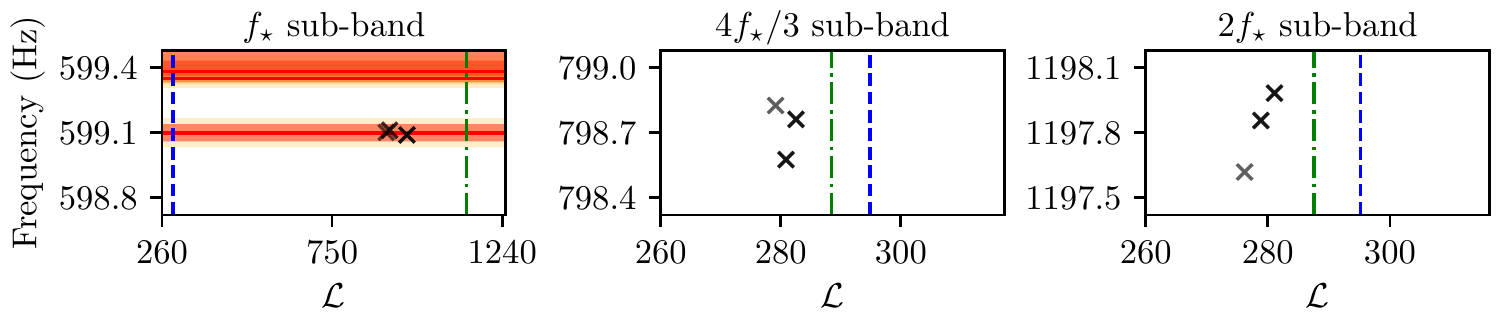}
		\caption{Search results for \igra.}
		\label{fig:igra}	
	\end{subfigure}
	\par\vspace{2em} %
	\begin{subfigure}[t]{0.8\textwidth}
		\includegraphics[width=\linewidth]{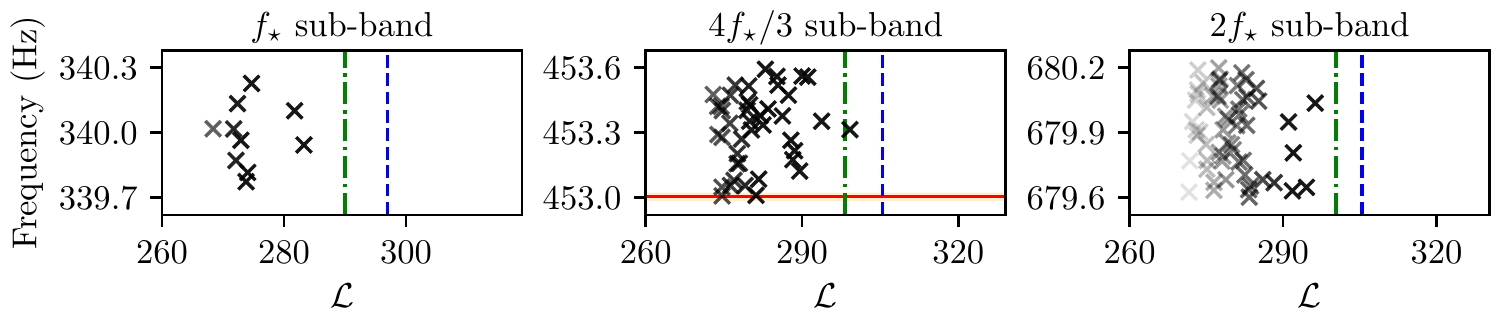}
		\caption{Search results for \maxi.}
		\label{fig:maxi}	
	\end{subfigure}
	\par\vspace{2em} %
	\begin{subfigure}[t]{0.8\textwidth}
		\includegraphics[width=\linewidth]{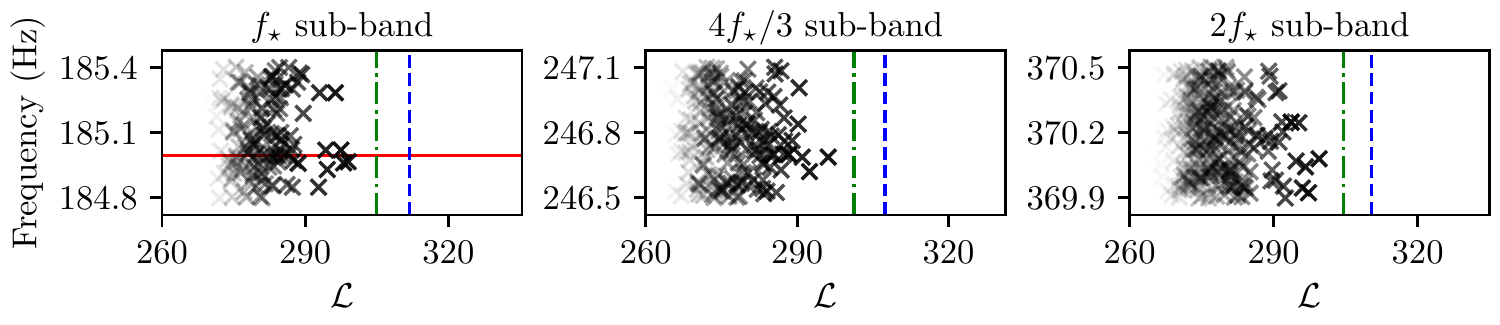}
		\caption{Search results for \xtea.}
		\label{fig:xtea}	
	\end{subfigure}
	\par\vspace{2em} %	
	\begin{subfigure}[t]{0.8\textwidth}
		\includegraphics[width=\linewidth]{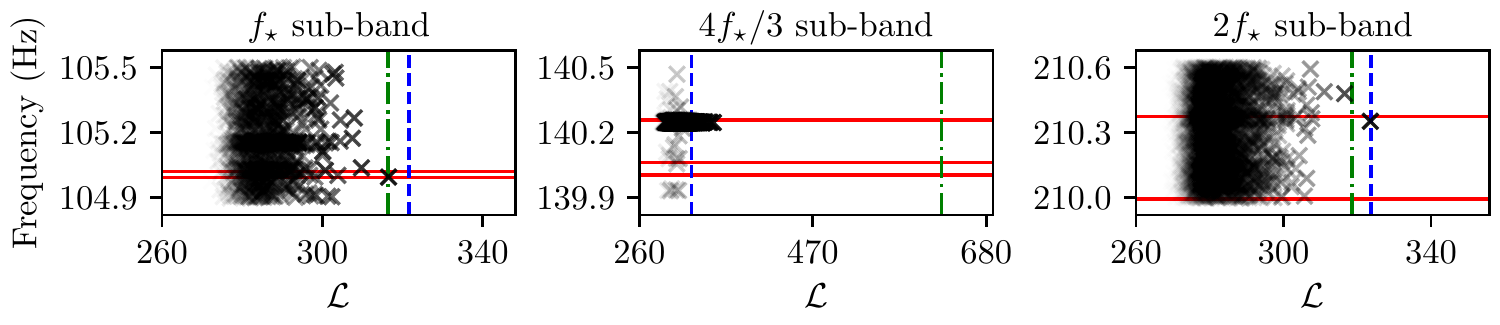}
		\caption{Search results for \igrb.}
		\label{fig:igrb}	
	\end{subfigure}
	\par\vspace{2em} %
	\begin{subfigure}[t]{0.8\textwidth}
		\includegraphics[width=\linewidth]{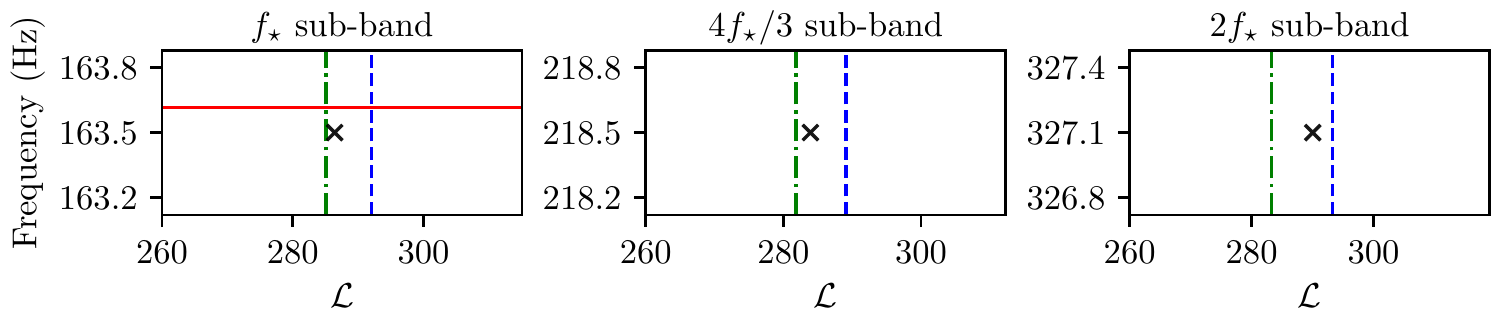}
		\caption{Search results for \igrc.}
		\label{fig:igrc}	
	\end{subfigure}
\end{figure*}
\begin{figure*}
	\ContinuedFloat
	\centering
	\begin{subfigure}[t]{0.8\textwidth}
		\includegraphics[width=\linewidth]{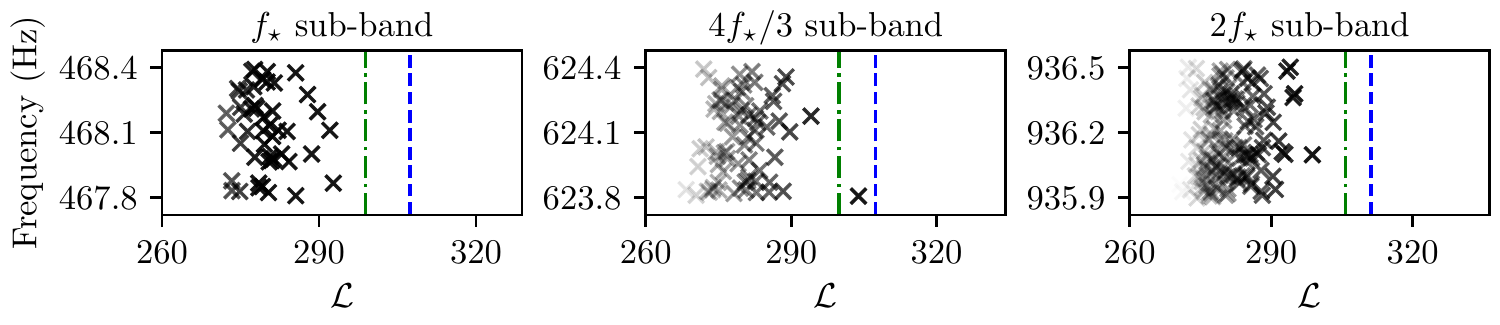}
		\caption{Search results for \igrd.}
		\label{fig:igrd}	
	\end{subfigure}
	\par\vspace{2em} %
	\begin{subfigure}[t]{0.8\textwidth}
		\includegraphics[width=\linewidth]{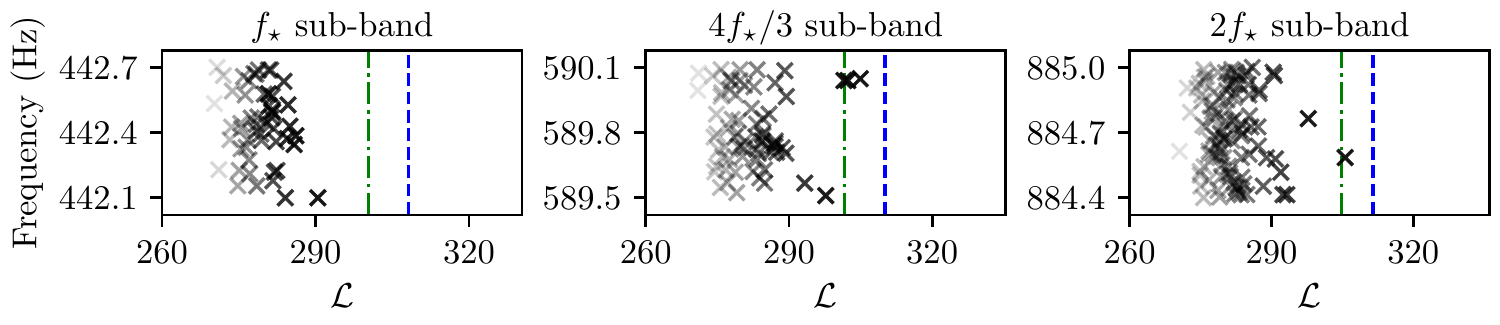}
		\caption{Search results for \saxb.}
		\label{fig:saxb}	
	\end{subfigure}
	\par\vspace{2em} %
	\begin{subfigure}[t]{0.8\textwidth}
		\includegraphics[width=\linewidth]{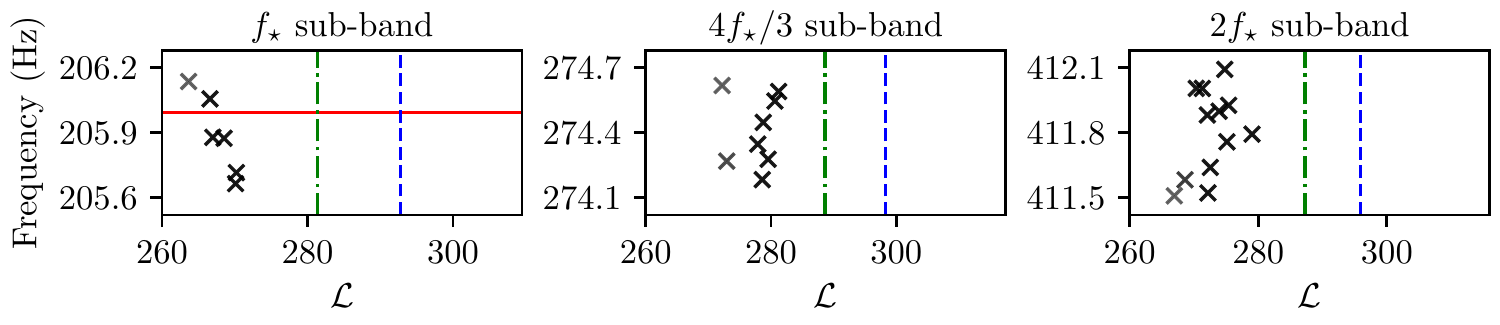}
		\caption{Search results for \ngc.}
		\label{fig:ngc}	
	\end{subfigure}
	\par\vspace{2em} %
	\begin{subfigure}[t]{0.8\textwidth}
		\includegraphics[width=\linewidth]{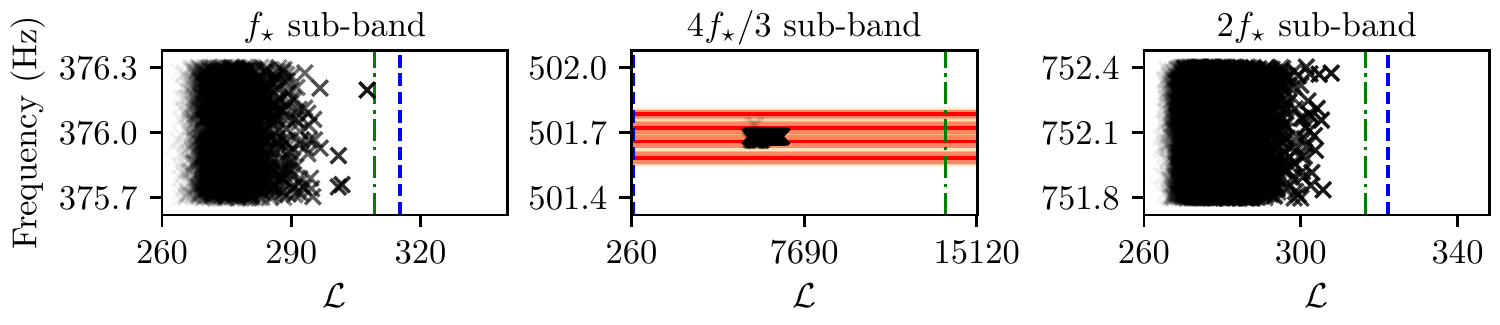}
		\caption{Search results for \igri.}
		\label{fig:igri}	
	\end{subfigure}
	\par\vspace{2em} %
	\begin{subfigure}[t]{0.8\textwidth}
		\includegraphics[width=\linewidth]{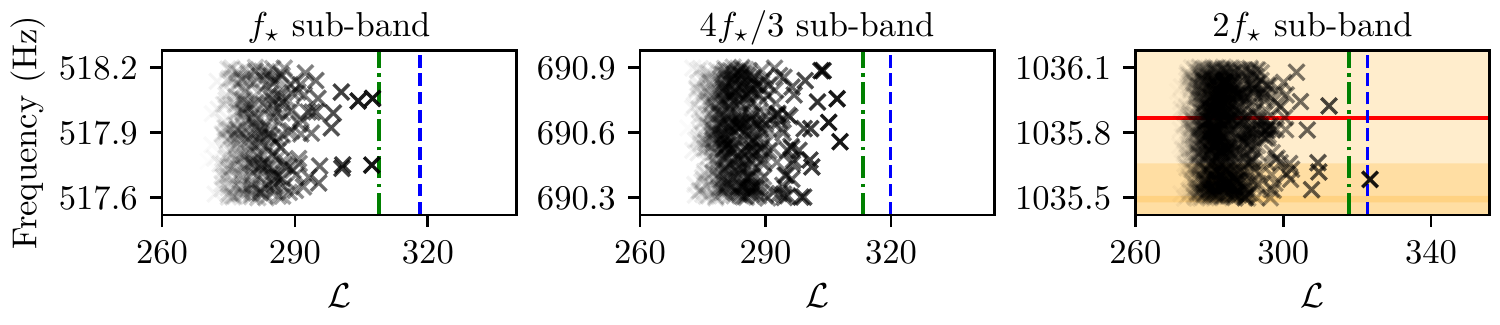}
		\caption{Search results for \swiftb.}
		\label{fig:swiftb}	
	\end{subfigure}
\end{figure*}
\begin{figure*}
	\ContinuedFloat
	\centering
	\begin{subfigure}[t]{0.8\textwidth}
		\includegraphics[width=\linewidth]{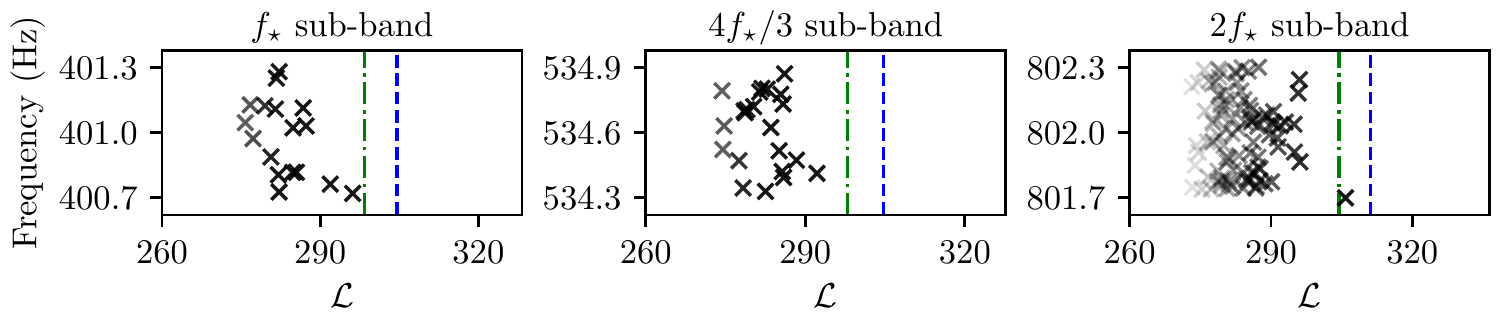}
		\caption{Search results for \igre.}
		\label{fig:igre}	
	\end{subfigure}
	\par\vspace{2em} %
	\begin{subfigure}[t]{0.8\textwidth}
		\includegraphics[width=\linewidth]{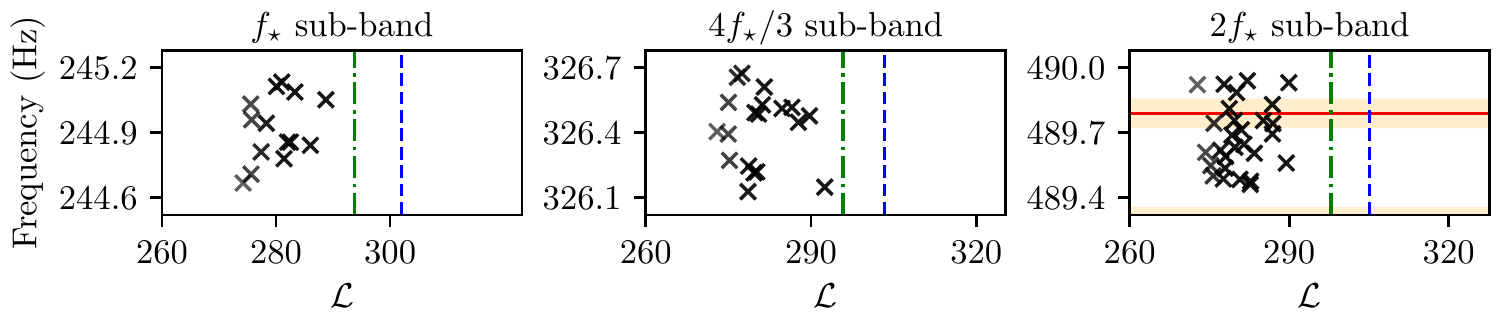}
		\caption{Search results for \igrf.}
		\label{fig:igrf}	
	\end{subfigure}
	\par\vspace{2em} %
	\begin{subfigure}[t]{0.8\textwidth}
		\includegraphics[width=\linewidth]{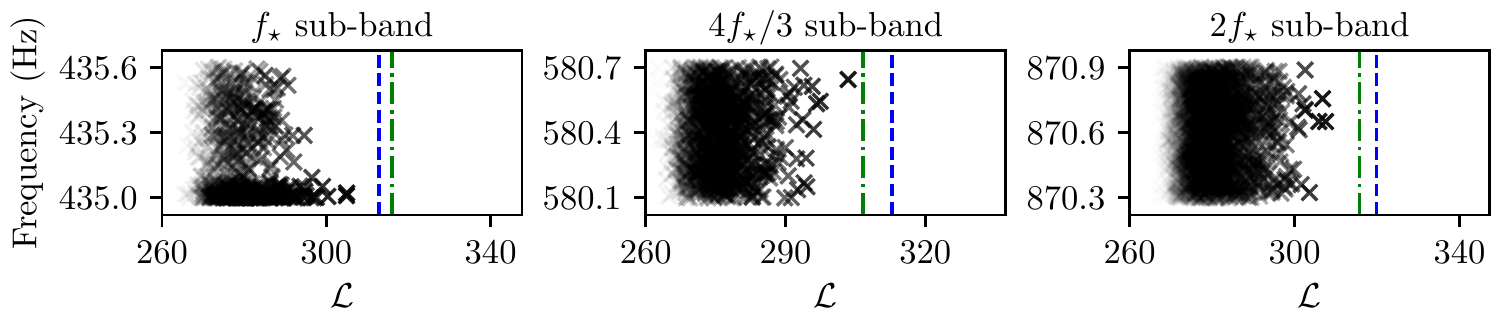}
		\caption{Search results for \xteb.}
		\label{fig:xteb}	
	\end{subfigure}
	\par\vspace{2em} %
	\begin{subfigure}[t]{0.8\textwidth}
		\includegraphics[width=\linewidth]{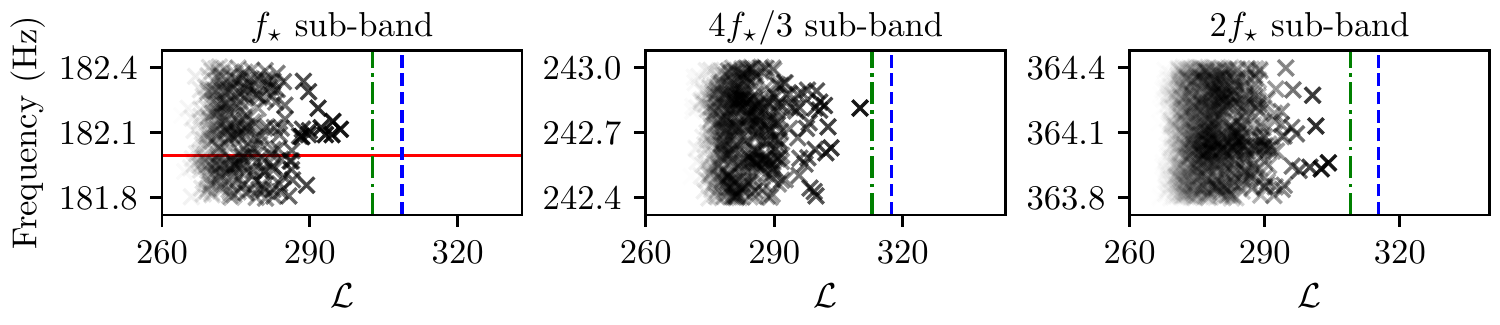}
		\caption{Search results for \swift.}
		\label{fig:swift}	
	\end{subfigure}
	\par\vspace{2em} %
	\begin{subfigure}[t]{0.8\textwidth}
		\includegraphics[width=\linewidth]{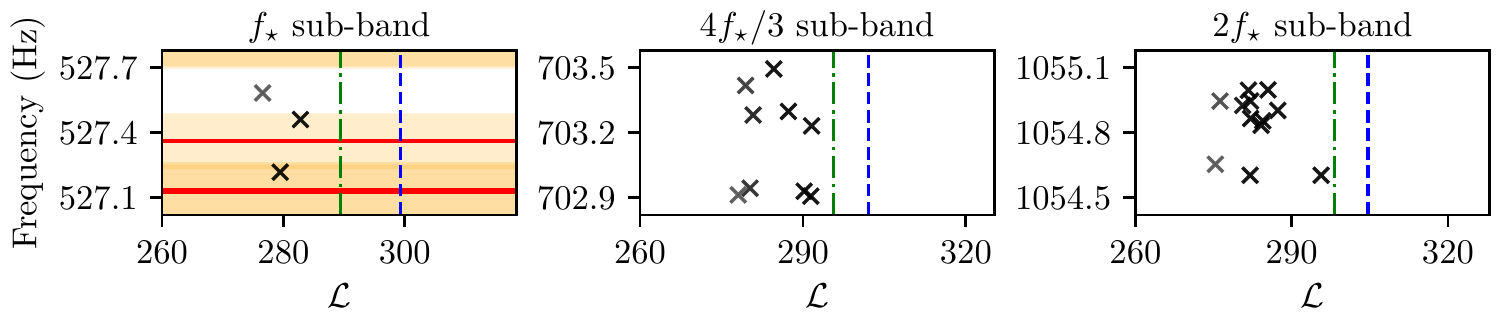}
		\caption{Search results for \igrg.}
		\label{fig:igrg}	
	\end{subfigure}
\end{figure*}
\begin{figure*}
	\ContinuedFloat
	\centering			
	\begin{subfigure}[t]{0.8\textwidth}
		\includegraphics[width=\linewidth]{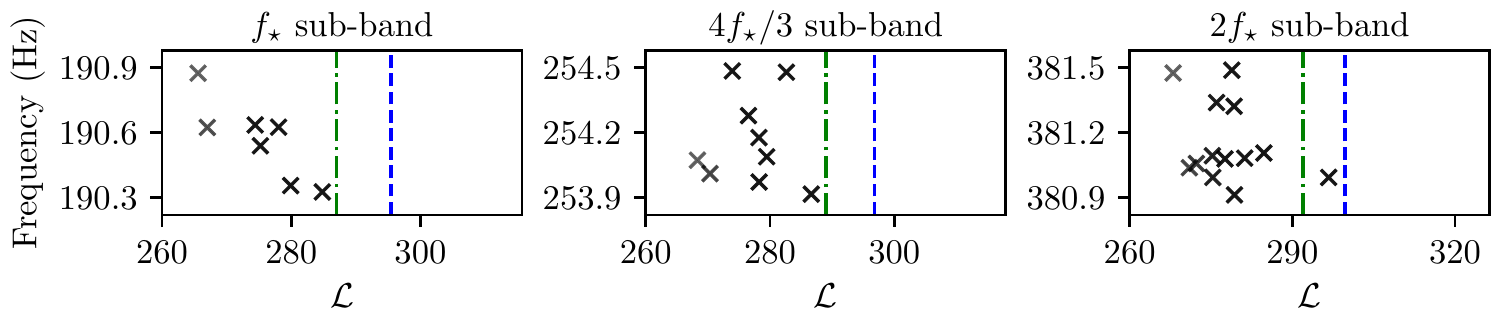}
		\caption{Search results for \xtec.}
		\label{fig:xtec}	
	\end{subfigure}
	\par\vspace{2em} %
	\begin{subfigure}[t]{0.8\textwidth}
		\includegraphics[width=\linewidth]{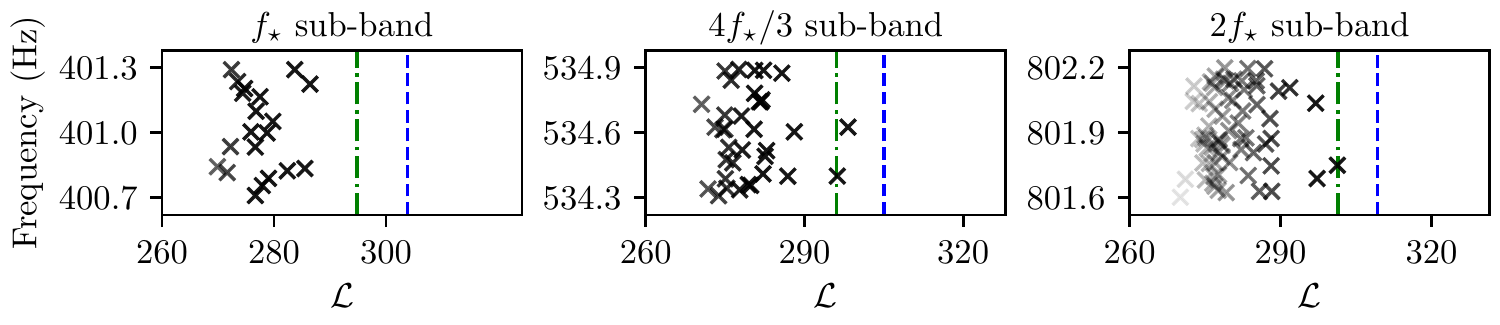}
		\caption{Search results for \sax.}
		\label{fig:sax}	
	\end{subfigure}
	\par\vspace{2em} %
	\begin{subfigure}[t]{0.8\textwidth}
		\includegraphics[width=\linewidth]{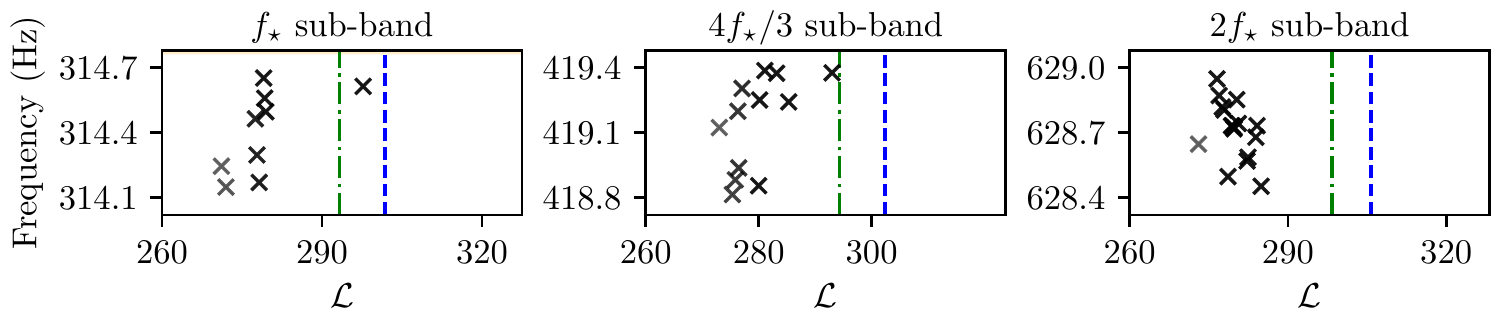}
		\caption{Search results for \xted.}
		\label{fig:xted}	
	\end{subfigure}
	\par\vspace{2em} %
	\begin{subfigure}[t]{0.8\textwidth}
		\includegraphics[width=\linewidth]{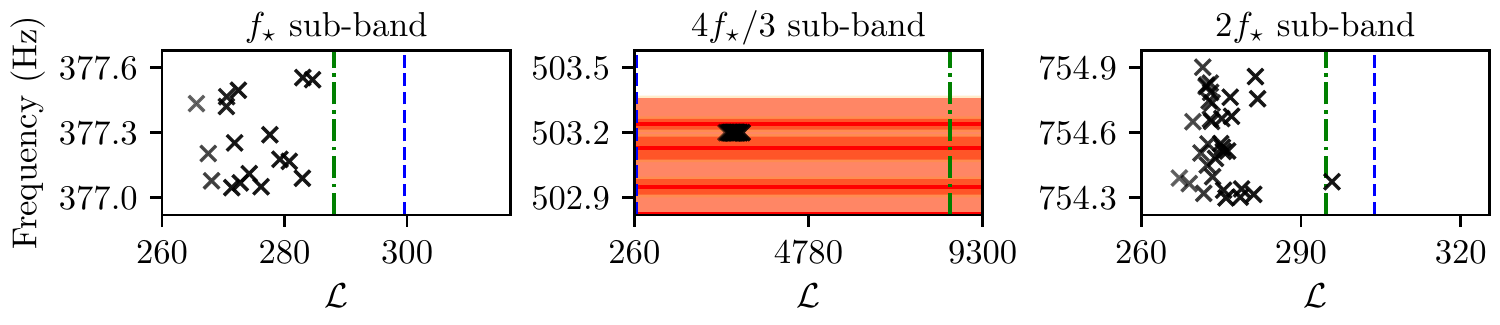}
		\caption{Search results for \hete.}
		\label{fig:hete}	
	\end{subfigure}
	\par\vspace{2em} %
	\caption{Search results for each target and sub-band, laid out as in Fig \ref{fig:igrh}. Black crosses indicate the frequency and $\ml$ for the most likely path through the sub-band for each binary template. The vertical blue dashed (green dot-dashed) lines correspond to the threshold set via Gaussian (off-target) noise realizations, $\lthg$ ($\lthot$), in each sub-band. Solid red lines indicate the peak frequency of known instrumental lines in the Hanford or Livingston detectors; the red band indicates the width of the line and the yellow band indicates the increased effective width due to Doppler broadening, as described in Sec.~\ref{sec:line_veto}. Multiple overlapping orange bands creates the red bands. The transparency of crosses in sub-bands with many templates, e.g.~the sub-bands of \igrb, is adjusted relative to the maximum $\ml$ in that sub-band for clarity.}
\par\vspace{5em}
\label{fig:allfull}
\end{figure*}

\begin{figure*}
	\centering
	\begin{subfigure}[t]{0.48\textwidth}
		\includegraphics[width=\linewidth]{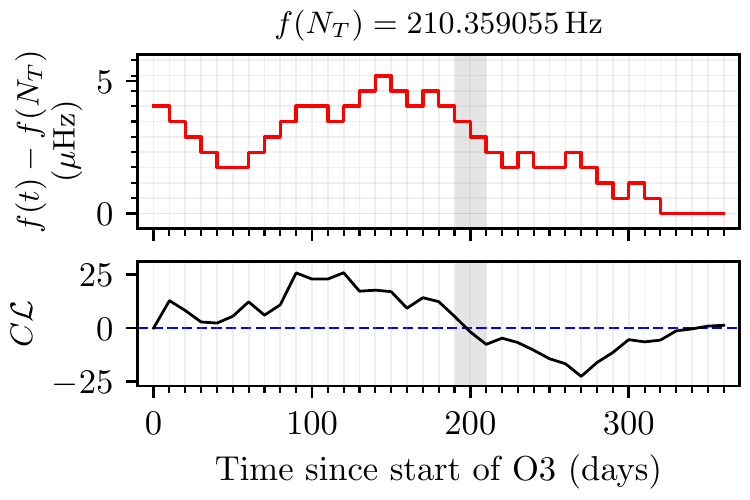}
		\caption{Surviving candidate 2 from \igrb.}
		\label{fig:igrb_f}	
	\end{subfigure}
	\hfill
	\begin{subfigure}[t]{0.48\textwidth}
		\includegraphics[width=\linewidth]{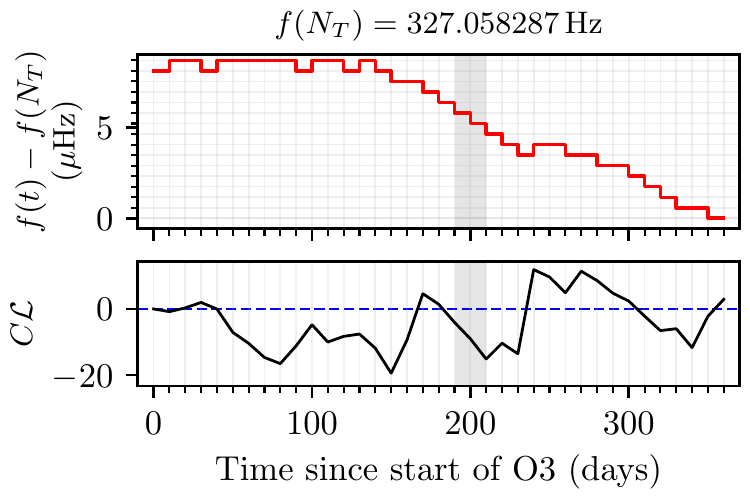}
		\caption{Surviving candidate 3 from \igrc.}
		\label{fig:igrc_f}	
	\end{subfigure}
	\par\medskip %
	\begin{subfigure}[t]{0.48\textwidth}
		\includegraphics[width=\linewidth]{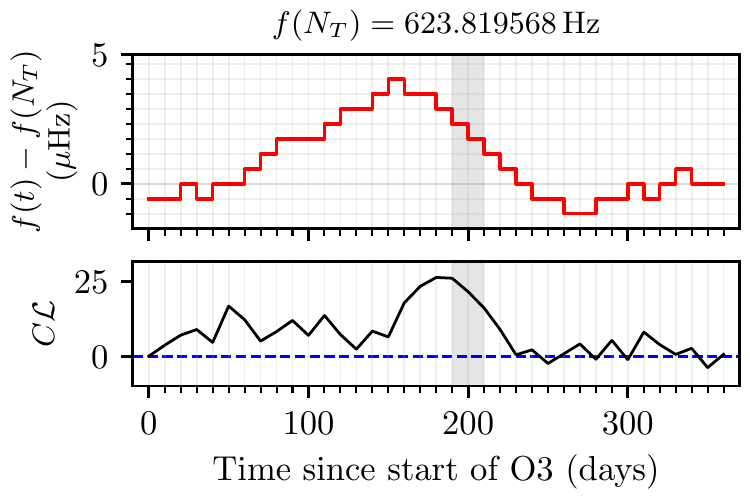}
		\caption{Surviving candidate 1 from \igrd.}
		\label{fig:igrd_f}	
	\end{subfigure}
	\hfill
	\begin{subfigure}[t]{0.48\textwidth}
		\includegraphics[width=\linewidth]{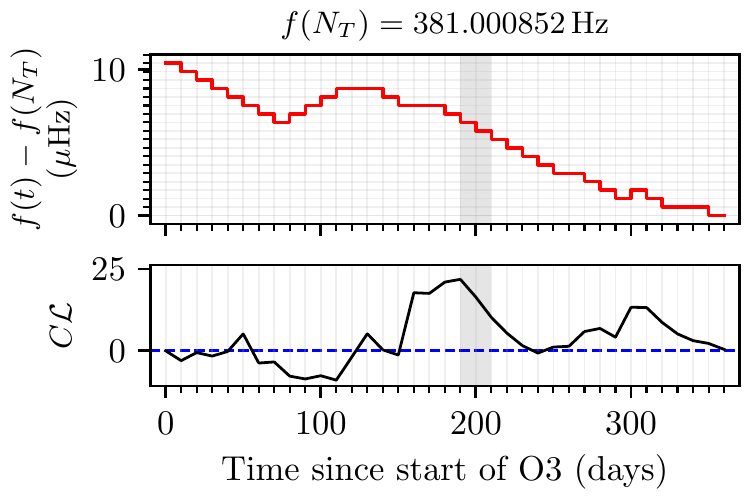}
		\caption{Surviving candidate 1 from \xtec.}
		\label{fig:xtec_f}
	\end{subfigure}
	\par\medskip %
	\begin{subfigure}[t]{0.48\textwidth}
		\includegraphics[width=\linewidth]{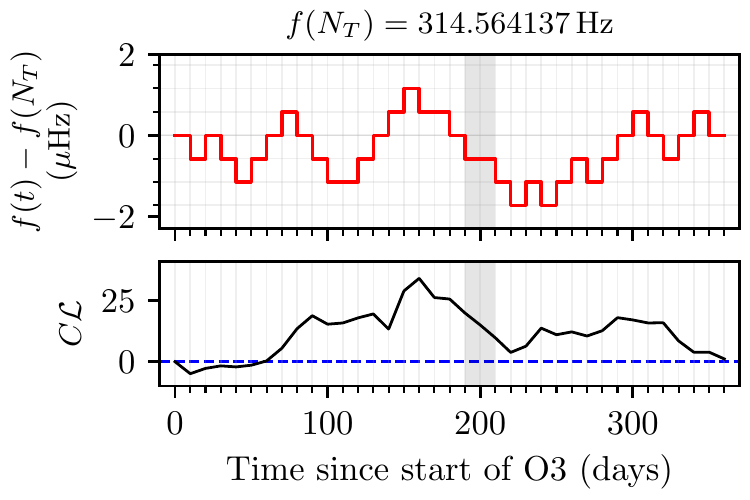}
		\caption{Surviving candidate 1 from \xted.}
		\label{fig:xted_f}
	\end{subfigure}
	\caption{Top panels: frequency paths, $f(t)$, for candidates with $\pn \leq 0.1$. The terminating frequency bin, $f(N_T)$, is subtracted and displayed in the title of each figure for clarity. Faint horizontal grey lines demarcate frequency bins of size $\Delta f = 5.787037\times10^{-7}\,$Hz, while faint vertical grey lines demarcate chunks of length $\tdr = 10\,$d. Bottom panels: the cumulative log-likelihood along the frequency path relative to the average sum log-likelihood needed to reach $\lth$, $C\ml \equiv \sum_{i=0}^{i=t}\big[\ml(i) - \lth / N_T\big]$, where $\sum_{i=0}^{i=t}\ml(i)$ is ${\rm ln\,} P(Q^*|O)$ from Eq.~\eqref{eq:probq_o} truncated after the $t$-th segment. The horizontal blue dashed line corresponds to $\sum_{i=0}^{i=t}\ml(i) = t \lth / N_T$. The grey shaded regions in both top and bottom panels correspond to the segments which have no SFTs and are therefore filled with a uniform log-likelihood, as described in Sec.~\ref{sec:o3data}.}
\label{fig:followups}
\end{figure*}

\begin{figure}
\centering
\includegraphics[width=\linewidth]{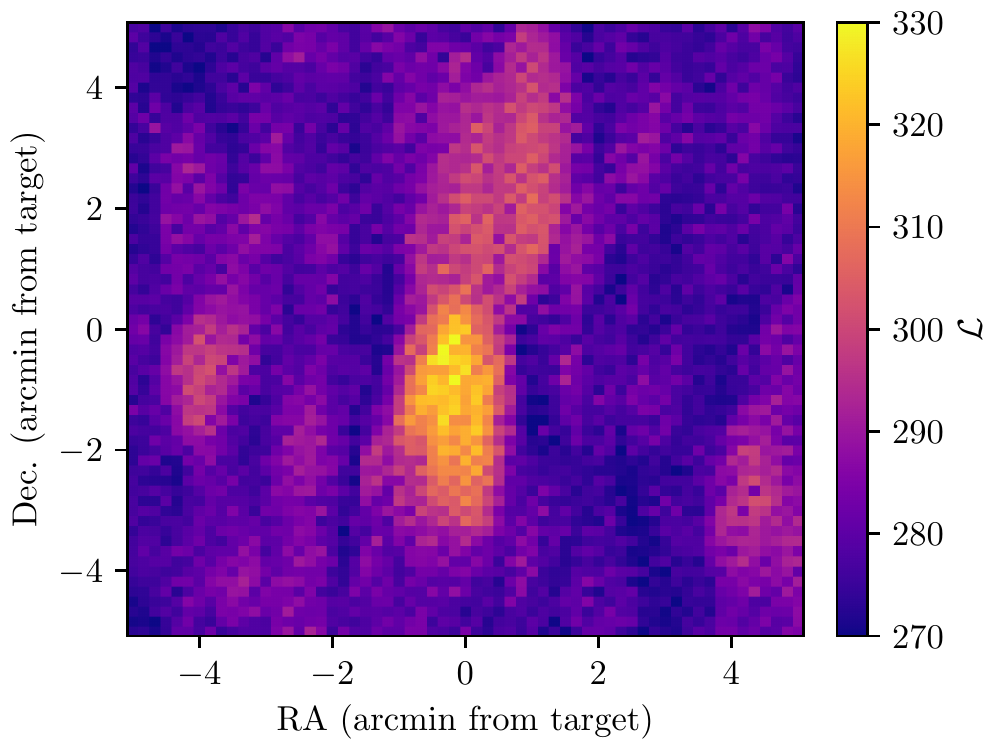}
\caption{$\ml$, as represented by the color of each pixel, calculated at 3721 regularly spaced sky locations in a 100\,arcmin$^2$ patch of sky, centered on \igrb. See text in Appendix \ref{app:followup} for details.}
\label{fig:skymap}
\end{figure}

\section{Survivor follow-up for target-of-opportunity search candidate \label{app:shortsax_followup}}
For posterity, and to aid future follow-up with different pipelines, we record in Table \ref{tab:shortsaxoutliers} the template, the frequency $f$, the log-likelihood $\ml$, and $\pn$, of the candidate from the target-of-opportunity search in Sec.~\ref{sec:shortsax} that survives all vetoes. 

\begin{table*}
\caption{Orbital template, $(P,\ a_0,\ \tasc)$, frequency, $f$, log-likelihood, $\ml$, and the probability of seeing a candidate at least this loud in pure noise, $\pn$, for the remaining candidate from the target-of-opportunity, $24\,$d coherent search when \sax\ was in outburst. The candidate cannot be eliminated by any of the vetoes detailed in Sec.~\ref{sec:vetoes}. \label{tab:shortsaxoutliers}}
\begin{ruledtabular}
\begin{tabular}{r r r r r r}
\textrm{$P$ (s)} & \textrm{$a_0$ (lt-s)} & \textrm{$\tasc$ (GPS time)} & \textrm{$f$ (Hz)} & \textrm{$\ml$} & \textrm{$\pn$} \\
\midrule
7249.155 & 0.062809 & 1249163578.03125 & 400.59656098 & 42.5 & 0.02 \\
\end{tabular}
\end{ruledtabular}
\end{table*}

As in Appendix \ref{app:followup}, we perform additional follow-up for this remaining candidate. With $T_{\rm obs}=24\,$d the point-spread-function of a moderately loud injection ($\ml \gtrsim \lth + 20$), at the sky location of the target, is a narrow ellipse $\sim2\,$arcmin wide in RA, but over $\sim30\,$arcmin tall in Dec. When we search a 100\,arcmin$^2$ patch of sky around the location of \sax\ we do not see any evidence of this point-spread-function at the source location. There is an ellipse with $\ml > \lth$ roughly $-2\,$arcmin away in RA from \sax, but as the location of the target is known to sub-arcsec precision \cite{Bult2020}, this ellipse is likely a noise fluctuation, rather than an astrophysical signal.

We also calculate $\ml$ in a small, densely sampled patch of the $\{P,\ \tasc\}$ parameter space around the candidate's template. As discussed in Appendix \ref{app:followup}, moderately loud injections ($\ml \gtrsim \lth + 20$) ``spread out'' in the $\{P,\ \tasc\}$ plane. However, the candidate is not loud enough for this diagnostic to provide evidence for or against the hypothesis that the candidate is a noise fluctuation.

If we assume that the remaining candidate is a false alarm, we calculate $\hul$ for the $24\,$d coherent search, using the procedure outlined in Sec.~\ref{sec:ul_method}. We find $\hul = 1.3\times 10^{-25}$ for the sub-bands centered on $f_\star$ and $4f_\star/3$, and $\hul = 1.7\times 10^{-25}$ for the sub-band centered on $2f_\star$. These upper limits are higher than the ones listed in Sec.~\ref{sec:ul_results} because the longer coherence time does not completely compensate for the shorter observation time.

Finally, we perform a complementary follow-up search using a deterministic signal template on the candidate of interest using \texttt{PyFstat} \cite{Ashton:2018ure, Keitel:2021xeq}. The use of the \texttt{PyFstat} algorithm as a follow-up technique was applied to the last surviving outlier of Ref.~\cite{o3aAllSkyIso} and previously in Refs.~\cite{Covas:2020nwy, o3abinaryallsky}.
The follow-up procedure, thoroughly described in Ref.~\cite{Tenorio:2021njf}, uses a Markov chain Monte Carlo (MCMC) sampler \cite{emcee, ptemcee} to explore a parameter-space region using the $\mathcal{F}$-statistic 
as log-likelihood \cite{Jaranowski1998}. Two coherence times are used here, namely $T_{\textrm{coh}} = 12\,\textrm{d}$ and $T_{\textrm{coh}} = 24\,\textrm{d}$. Prior distributions 
are Gaussian distributions centered at the outlier parameters (Table~\ref{tab:shortsaxoutliers}) using a standard deviation of one parameter-space bin with maximum mismatch $\mu_{\textrm{max}}=1$ \cite{Leaci2015}. 
The results of the follow-up are evaluated using a Bayes factor, $\mathcal{B}_{\textrm{S}/\textrm{N}}$, that compares the evidence for a model that the data contain a coherent signal to the 
evidence for a model that the data contain only noise.
The value of $\mathcal{B}_{\textrm{S}/\textrm{N}}$ is computed by comparing the change in the $\mathcal{F}$-statistic of the 
loudest candidate between the two follow-up stages with different coherence times: if a signal is present in the data, the $\mathcal{F}$-statistic should provide a consistent estimate of the  
signal-to-noise ratio; otherwise, the loudest candidate is a result of noise, the distribution of which follows a Gumbel distribution. This noise distribution is estimated using a similar method to the one described in Appendix~\ref{app:offt_th}, with 600 off-source calculations performed. 

The loudest candidate of the follow-up returns a log-Bayes factor of 
$\log_{10}\mathcal{B}_{\textrm{S}/\textrm{N}} = 1.45$. We characterize the 
$\log_{10}\mathcal{B}_{\textrm{S}/\textrm{N}}$ distribution using 400 isotropically distributed sources
injected into the real data with an amplitude of $\hul$. We obtain a 1\% false dismissal threshold of 8.75, 
which is significantly larger than the candidate's log-Bayes factor of 1.45. That is, if this were a true signal, 
with $h_0 = \hul$, we would expect the log-Bayes factor to be higher than what we see in the real data by about 7.
We conclude that there is no significant evidence of continuous gravitational wave emission from this target.

\bibliography{CW_AMXP_O3_bib}

\end{document}

%% file: authors.tex
\author{R.~Abbott}
\affiliation{LIGO Laboratory, California Institute of Technology, Pasadena, CA 91125, USA}
\author{T.~D.~Abbott}
\affiliation{Louisiana State University, Baton Rouge, LA 70803, USA}
\author{F.~Acernese}
\affiliation{Dipartimento di Farmacia, Universit\`a di Salerno, I-84084 Fisciano, Salerno, Italy}
\affiliation{INFN, Sezione di Napoli, Complesso Universitario di Monte S. Angelo, I-80126 Napoli, Italy}
\author{K.~Ackley}
\affiliation{OzGrav, School of Physics \& Astronomy, Monash University, Clayton 3800, Victoria, Australia}
\author{C.~Adams}
\affiliation{LIGO Livingston Observatory, Livingston, LA 70754, USA}
\author{N.~Adhikari}
\affiliation{University of Wisconsin-Milwaukee, Milwaukee, WI 53201, USA}
\author{R.~X.~Adhikari}
\affiliation{LIGO Laboratory, California Institute of Technology, Pasadena, CA 91125, USA}
\author{V.~B.~Adya}
\affiliation{OzGrav, Australian National University, Canberra, Australian Capital Territory 0200, Australia}
\author{C.~Affeldt}
\affiliation{Max Planck Institute for Gravitational Physics (Albert Einstein Institute), D-30167 Hannover, Germany}
\affiliation{Leibniz Universit\"at Hannover, D-30167 Hannover, Germany}
\author{D.~Agarwal}
\affiliation{Inter-University Centre for Astronomy and Astrophysics, Pune 411007, India}
% ...
